\documentclass[12pt]{article}
\usepackage[paperwidth=21cm,paperheight=26cm, textwidth=16cm,textheight=20.6cm]{geometry}
\usepackage{amsmath,amsthm,amssymb,xspace,xcolor,hyperref}\hypersetup{colorlinks=true, linkcolor=black,citecolor=black,urlcolor=black}\usepackage{graphicx}
\numberwithin{equation}{section}

\theoremstyle{plain}
\newtheorem{theorem}{Theorem}

\newtheorem{claim}{Claim}[section]
\newtheorem{proposition}{Proposition}
\newtheorem{lemma}{Lemma}
\newtheorem{corollary}{Corollary}
\newtheorem{definition}{Definition}
\newtheorem{question}{Question}
\newtheorem{hypothesis}{Hypothesis}

\newcommand{\APC}{\ensuremath{{\sf APC_1}}\xspace}

\newcommand{\SB}{\ensuremath{{\sf S^1_2}}\xspace}

\newcommand{\PV}{\ensuremath{{\sf PV_1}}\xspace}

\newcommand{\Ptime}{\ensuremath{{\sf P}}\xspace}
\newcommand{\Ppoly}{\ensuremath{{\sf P/poly}}\xspace}
\newcommand{\NC}{\ensuremath{{\sf NC}^1}\xspace}
\newcommand{\AC}{\ensuremath{{\sf AC}^0}\xspace}
\newcommand{\NPtime}{\ensuremath{{\sf NP}}\xspace}
\newcommand{\NP}{\ensuremath{{\sf NP}}\xspace}
\newcommand{\coNP}{\ensuremath{{\sf coNP}}\xspace}

\newcommand{\Circuit}{\ensuremath{{\sf Circuit}}\xspace}

\newcommand{\SAT}{\ensuremath{{\sf SAT}}\xspace}
\newcommand{\PARITY}{\ensuremath{{\sf PARITY}}\xspace}
\newcommand{\MCSP}{\ensuremath{{\sf MCSP}}\xspace}
\newcommand{\GCSP}{\ensuremath{{\sf GCSP}}\xspace}

\newcommand{\LB}{\ensuremath{{\sf LB}}\xspace}

\newcommand{\ttable}{\ensuremath{{\sf tt}}\xspace}
\newcommand{\lb}{\ensuremath{{\sf lb}}\xspace}

\begin{document}

\title{Learning algorithms from circuit lower bounds}
\author{J\'an Pich \\ \small University of Oxford} 
\date{November 2020}
\maketitle

\begin{abstract} 
We revisit known constructions of 
efficient learning algorithms from various notions of constructive circuit lower bounds such as distinguishers breaking pseudorandom generators or efficient witnessing algorithms which find errors of small circuits attempting to compute hard functions. 
As our main result we prove 
that if it is possible to find efficiently, in a particular interactive way, errors of many p-size circuits attempting to solve hard problems, then p-size circuits can be PAC learned over the uniform distribution with membership queries by circuits of subexponential size. The opposite implication holds as well. This provides a new characterisation of learning algorithms and extends the natural proofs barrier of Razborov and Rudich. The proof is based on a method of exploiting Nisan-Wigderson generators introduced by Kraj\'{i}\v{c}ek (2010) and used to analyze complexity of circuit lower bounds in bounded arithmetic.

An interesting consequence of known constructions of learning algorithms from circuit lower bounds is 
a learning speedup of Oliveira and Santhanam (2016). 
We present an alternative proof of this phenomenon and discuss its potential to advance the program of hardness magnification. 
\end{abstract}


\section{Introduction}

While the central conjectures in complexity theory such as $\Ptime\ne\NPtime$ have the form of impossibility results, we hope that a better understanding of the impossibility phenomena will also shed light on the question of constructing new useful algorithms. A successful formalization of such hopes can be found in cryptography, where the impossibility results in the form of average-case lower bounds are turned into cryptographic primitives. In the present paper we are interested in turning complexity lower bounds into efficient learning algorithms.  
\medskip

Results of this form can be traced back to cryptography as well. The 
`{\em pseudorandomness from unpredictability}' paradigm was used by Blum, Furst, Kearns and Lipton~\cite{BFKL} to show that efficient distinguishers breaking pseudorandom generators 
imply an efficient learning of p-size circuits on average. The distinguishers from \cite{BFKL} can be interpreted as constructive circuit lower bounds distinguishing partial truth-tables of easy Boolean functions from partial truth-tables of hard functions, cf. Section \ref{s:leargen}. The existing methods for proving circuit lower bounds have been also applied in constructions of new learning algorithms for restricted circuit classes, e.g. Linial, Mansour and Nisan \cite{LMN} used \AC lower bounds to get learning algorithms for \AC. More recently, in a landmark work, Carmosino, Impagliazzo, Kabanets and Kolokolova \cite{CIKK} gave a generic construction of learning algorithms from natural proofs of circuit lower bounds. Oliveira and Santhanam \cite{OS} extended their result to a dichotomy between the non-existence of non-uniform pseudorandom function families and the existence of efficient learning of small circuits. These results led Oliveira and Santhanam \cite{OS} also to a discovery of a surprising learning speedup. For example, learning p-size circuits over the uniform distribution with membership queries by circuits of weakly subexponential size $2^n/n^{\omega(1)}$ implies that for each constant $k$ and $\epsilon>0$, circuits of size $n^k$ can be learned over the uniform distribution with membership queries by circuits of strongly subexponential size $2^{n^{\epsilon}}$.


\subsection{Our contribution}

In the present paper we revisit these connections. We start by considering a simple {\em instance-specific} model of learning in which proving a single circuit lower bound implies a reliable prediction of the value of a target function on a single input. The model underlies the construction of learning algorithms from \cite{BFKL, CIKK} and differs from the standard PAC learning model mainly in that it does not ask learners to construct a circuit which computes the target function on a big fraction of inputs, cf. Section \ref{s:i-s}. 

\medskip
\noindent {\bf Learning from witnessing lower bounds.} Our main result is a construction of efficient PAC learning of p-size circuits from a constructive circuit lower bound for an arbitrary Boolean function $H$. More precisely, we obtain subexponential-size circuits learning p-size circuits over the uniform distribution with membership queries. The assumption of a constructive circuit lower bound we need is defined as the existence of $2^{O(n)}$-size `witnessing' circuits $W$ which given an oracle access to a p-size circuit $D$ with $n$ inputs find a not-yet-queried input on which $D$ fails to compute $H$. The circuits $W$ are allowed to fail on $1/poly(n)$ fraction of circuits $D$. 
Moreover, even if circuits $W$ succeed on a circuit $D$ they are allowed to output incorrect answer $\log n$ times (receiving a correction in each round) before generating the right answer, cf. Theorem \ref{t:main}. The implication can be also interpreted as a construction of PAC learning algorithms from a frequent interactive instance-specific\footnote{We use the adjective `instance-specific' only informally in this paper. The instance-specific model discussed earlier actually differs slightly from the concept in Theorem \ref{t:main}.} learning: If we are given an algorithm which is able to predict a value of a big fraction of p-size circuits (after a small number of queries and $\le \log n$ mistakes) even on a single input, this already implies learnability of p-size circuits on almost all inputs. The opposite implication producing efficient witnessing of lower bounds from learning algorithms holds as well, which yields a new characterisation of PAC learning of small circuits, cf. Lemma~\ref{l:mainconverse}.

\medskip
\noindent {\bf {\em Relation to proof complexity, natural proofs and witnessing theorems.}} 
The notion of interactive witnessing of circuit lower bounds from Theorem \ref{t:main} is motivated by witnessing theorems from bounded arithmetic. One of the most prominent theories of bounded arithmetic is Cook's theory \PV, which formalizes p-time reasoning. Theories of bounded arithmetic satisfy many so called witnessing theorems, which allow us to show, for example, that if we can prove a p-size circuit lower bound for a function $H\in\NP$ in \PV then there exists a witnessing analogous to the one from Theorem~\ref{t:main} except that the witnessing circuits $W$ have white-box access to $D$ (i.e. access to a full description of $D$), see Section \ref{s:witnessing} for a more detailed comparison. The witnessing from Theorem \ref{t:main} is also closely related to algorithms finding hard instances of \NP problems by Gutfreund, Shaltiel, Ta-Shma \cite{GST} and Atserias \cite{Abb}. The main difference is that the algorithms from~\cite{GST} have white-box access to the algorithm whose error they search for. While Atserias \cite{Abb} made \cite{GST} work with the black-box (oracle) access, his algorithm achieves much smaller probability of success than the one required in Theorem \ref{t:main}, cf. Section \ref{s:witnessing}. 

The proof of Theorem \ref{t:main} is an adaptation of a method of exploiting Nisan-Wigderson generators introduced by Kraj\'{i}\v{c}ek \cite{Knw} in order to give a model-theoretic evidence for Razborov's conjecture in proof complexity. Razborov's conjecture \cite{Rkdnf} states a conditional hardness of deriving tautologies expressing the existence of an element outside of the range of a suitable NW-generator in strong proof systems. Kraj\'i\v{c}ek's result significantly strengthens a similar but much simpler proof of the validity of Razborov's conjecture for proof systems with feasible interpolation \cite{Pnw}. The method has been also used to show a conditional hardness of generating hard tautologies \cite{Kht}, a conditional unprovability of p-size circuit lower bounds for \SAT in theories of bounded arithmetic below Cook's theory \PV \cite{Pclba} and an unconditional unprovability of strong nondeterministic lower bounds in Je\v{r}\'abek's theory of approximate counting \APC \cite{PSapc}. We take advantage of its unique way of exploiting the NW generator: it gives us a reconstruction algorithm which after breaking the NW-generator in a particular interactive fashion allows us to approximately compute the function on which the generator is based. There are, however, technical issues with adapting this method in our context, e.g. unlike in bounded arithmetic our witnessing circuits can fail with a significant probability. Our main contribution is in finding the right notions which allow the arguments to go through (in both directions). 

A competing notion of constructive circuit lower bounds has been developed in the influential theory of natural proofs of Razborov and Rudich \cite{RR}, which explains why many of the existing lower bound methods cannot yield separations such as $\Ptime\ne \NP$. Natural proofs are known to be equivalent to the existence of efficient learning algorithms, cf. \cite{CIKK}. For example, \Ppoly-natural proofs useful against \Ppoly\footnote{\Ppoly-natural proofs useful against \Ppoly are defined as $2^{O(n)}$-size circuits with $2^n$ inputs accepting a $1/2^{O(n)}$-fraction of inputs and rejecting all inputs which represent truth-tables of Boolean functions on $n$ inputs computable by p-size circuits, cf. Definition \ref{d:natpr}.} are equivalent to subexponential-size circuits learning p-size circuits over the uniform distribution with membership queries. Furthermore, natural proofs have been used to derive unprovability results in proof complexity as well. Specifically, to derive unprovability of circuit lower bounds in proof systems with the feasible interpolation property, cf. \cite{Rba,Kdual}. Despite similar applications and motivations for defining these concepts, the relation between natural proofs and the witnessing method has not been clear. In fact, a priori the `static' definition of natural proofs appears to be quite orthogonal to the witnessing from Theorem \ref{t:main}. Theorem \ref{t:main} thus not only extends the scope of the natural proofs barrier by providing another equivalent characterisation which incorporates interactivity but also helps to clarify its relation to the witnessing method. 

\medskip
\noindent {\bf Learning speedup.} Our second 
contribution is a simple proof of a generalized learning speedup 
of Oliveira and Santhanam \cite{OS}. 
Specifically, we show that for each superpolynomial function $s$, if for each constant $k$, circuits of size $n^k$ are learnable by circuits of size $s$ over the uniform distribution with random examples, then for each constant $k$ and $\epsilon>0$, circuits of size $n^k$ are learnable over the uniform distribution with membership queries by circuits of size $O(s^{\epsilon})$, cf. Theorem \ref{t:speedup}. 
We obtain the speedup by a more direct exploitation of a slightly modified NW-generator. In comparison to the proof from \cite{OS}, this sidesteps the need to construct natural proofs and invoke the construction of Carmosino et al. \cite{CIKK}. A disadvantage of the method is that we need to assume learning with random examples instead of membership queries. Nevertheless, we present one more alternative proof of the learning speedup based on (a simple case of) Theorem~\ref{t:main}, which allows to start with membership queries, cf. Theorem \ref{t:aspeed}. We emphasize, however, that behind all proofs of the learning speedup is essentially the same general idea of reconstructing, in this or that way, the base function of some form of the NW-generator.

\medskip
\noindent {\bf {\em Relation to hardness magnification and locality.}} The generalized learning speedup can be interpreted as a nonlocalizable hardness magnification theorem reducing a complexity lower bound into a seemingly weaker one. In general, hardness magnification refers to an approach to strong complexity lower bounds developed in a series of recent papers, cf. Section \ref{s:speedup}. Unfortunately, while the approach avoids (in certain cases provably \cite{CHOPRS}) the natural proofs barrier, it suffers from a `{\em locality barrier}': magnification theorems typically yield unconditional upper bounds for specific problems if the computational model in question is allowed to use oracles with small fan-in, but the existing lower bounds actually work even against the presence of local oracles. In fact, a better understanding of nonlocalizable lower bounds is essential for further progress on strong complexity lower bounds in general, see Section \ref{s:speedup} for more details. A promising aspect of the learning speedup (Theorem \ref{t:speedup}) is that it avoids the locality barrier, cf. Section \ref{s:speedup}.

\bigskip

\noindent {\bf Learning from breaking cryptographic pseudorandom generators.} In Section~\ref{s:leargen} we survey known constructions of learning algorithms from distinguishers breaking pseudorandom generators (PRGs) or natural proofs. 
While several such constructions are known, the question of extracting efficient learning of p-size circuits from the non-existence of cryptographic PRGs remains open. A positive answer to this question would establish an interesting win-win situation: either safe cryptography or efficient learning is possible. In the already mentioned approach, Oliveira and Santhanam \cite{OS} showed that efficient learning of p-size circuits with membership queries follows from the non-existence of nonuniform pseudorandom function families. By a straightforward adaptation of the proof method behind their result we show that efficient learning of p-size circuits {\em with random examples} follows from the non-existence of succinct nonuniform pseudorandom function families, cf. Theorem \ref{t:towardscore}. Finally, we point out that the desired construction of learning algorithms from the non-existence of cryptographic PRGs is closely related to a question of Rudich about turning demibits to superbits, cf. Section \ref{s:rudich}.

\section{Preliminaries}

$[n]$ denotes $\{1,\dots,n\}$. $\Circuit[s]$ denotes fan-in two Boolean circuits of size at most $s$. The size of a circuit is the number of gates. A function $f:\{0,1\}^n\mapsto \{0,1\}$ is $\gamma$-approximated by a circuit $C$, if $\Pr_x[C(x)=f(x)]\ge\gamma$.

\begin{definition}[Natural property \cite{RR}]\label{d:natpr} Let $m=2^n$ and $s,d:\mathbb{N} \mapsto \mathbb{N}$. A sequence of circuits $\{C_{m}\}^{\infty}_{m=1}$ is a $\Circuit[s(m)]$-natural property useful against $\Circuit[d(n)]$ if
\begin{itemize}
\item[1.] {\em Constructivity.} $C_{m}$ has $m$ inputs and size $s(m)$,
\item[2.] {\em Largeness.} $\Pr_x[C_{m}(x)=1]\ge 1/m^{O(1)}$,
\item[3.] {\em Usefulness.} For each sufficiently big $m$, $C_{m}(x)=1$ implies that $x$ is a truth-table of a function on $n$ inputs which is not computable by circuits of size $d(n)$.
\end{itemize}
\end{definition}


\begin{definition}[Pseudorandom generator]\label{d:prg}
A function $g:\{0,1\}^n\mapsto\{0,1\}^{n+1}$ computable by p-size circuits is a pseudorandom generator safe against circuits of size $s(n)$, if for each circuit $D$ of size $s(n)$, $$\left|\Pr_{y\in\{0,1\}^{n+1}}[D(y)=1]-\Pr_{x\in\{0,1\}^n}[D(g(x))=1]\right|<\frac{1}{s(n)}.$$
\end{definition}

\begin{definition}[PAC learning]\label{d:lear}
A circuit class $\mathcal{C}$ is learnable over the uniform disribution by a circuit class $\mathcal{D}$ up to error $\epsilon$ with confidence $\delta$, if there are randomized oracle circuits $L^f$ from $\mathcal{D}$ such that for every Boolean function $f:\{0,1\}^n\mapsto\{0,1\}$ computable by a circuit from $\mathcal{C}$, when given oracle access to $f$, input $1^n$ and the internal randomness $w \in \{0,1\}^*$, $L^f$ outputs the description of a circuit satisfying 
  \begin{equation*}
    \Pr_w [ L^f(1^n, w) \text{ } (1-\epsilon) \text{-approximates } f ] \geq \delta.
  \end{equation*}

\noindent $L^f$ uses non-adaptive membership queries if the set of queries which $L^f$ makes to the oracle does not depend on the answers to previous queries. $L^f$ uses random examples if the set of queries which $L^f$ makes to the oracle is chosen uniformly at random. 
\end{definition}

In this paper, PAC learning always refers to learning over the uniform distribution. 
\bigskip

\noindent {\bf Boosting confidence and reducing error.} The confidence of the learner can be efficiently boosted in a standard way. Suppose an $s$-size circuit $L^f$ learns $f$ up to error $\epsilon$ with confidence $\delta$. We can then run $L^f$ $k$ times, test the output of $L^f$ from every run with $m$ new random queries and output the most accurate one. By Hoeffding's inequality, $m$ random queries fail to estimate the error $\epsilon$ of an output of $L^f$ up to $\gamma$ with probability at most $2/e^{2\gamma^2m}$. 
Therefore the resulting circuit of size $poly(s,m,k)$ learns $f$ up to error $\epsilon+\gamma$ with confidence at least $1-2k/e^{2\gamma^2m}-(1-\delta)^k\ge 1-2k/e^{2\gamma^2m}-e^{-k\delta}$. If we are trying to learn small circuits  we can get even confidence 1 by fixing internal randomness of learner nonuniformly without losing much on the running time or the error of the output. It is also possible to reduce the error up to which $L^f$ learns $f$ without a significant blowup in the running time and confidence. If we want to learn $f$ with a better error, we first learn an amplified version of $f$, $Amp(f)$. Employing direct product theorems and Goldreich-Levin reconstruction algorithm, Carmosino et. al. \cite[Lemma 3.5]{CIKK} showed that for each $0<\epsilon,\gamma<1$ it is possible to map a Boolean function $f$ with $n$ inputs to a Boolean function $Amp(f)$ with $poly(n,1/\epsilon,\log(1/\gamma))$ inputs so that $Amp(f)\in \Ppoly^f$ and there is a probabilistic $poly(|C|,n,1/\epsilon,1/\gamma)$-time machine which given a circuit $C$ $(1/2+\gamma)$-approximating $Amp(f)$ and an oracle access to $f$ outputs with high probability a circuit $(1-\epsilon)$-approximating $f$. We thus typically ignore the optimisation of the confidence and error parameter in the rest of the paper.

\section{Instance-specific learning}\label{s:i-s}

The most direct way of turning circuit lower bounds into a certain type of learning can be described as follows.\footnote{The simple observation from box A appeared in \cite[Section 4.5]{MP} and \cite{Pmu15}. 
I am not aware of a more systematic treatment of this concept. There are related models of learning such as `knows what it knows' model by Li-Littman-Walsh \cite{LLW} and `reliable learning' by Rivest-Sloan \cite{RS} which prohibit incorrect predictions in various ways. These models, however, follow the formalization of PAC learning in that the goal of the learner is to learn the target concept by accessing it. In box A we do not assume that the target concept $f$ is determined on all inputs or prior to the given samples.}
\bigskip

\setlength\fboxrule{1pt}
\fbox{\parbox{411pt}{
\vspace{6pt}

\hspace{1pt} {\bf A. Prediction from lower bound.} Suppose we are given bits $f(y_1),\dots,f(y_k)$ for $n$-bit strings $y_1,\dots,y_k$ defining a partial Boolean function $f$. We want to predict the value of $f$ on a new input $y_{k+1}\in\{0,1\}^n$. A priori $f(y_{k+1})$ is not defined but we will interpret the minimal-size circuit $C^f$ coinciding with $f$ on $y_1,\dots,y_k$ as `the right' prediction of $f(y_{k+1})$. That is, we want to find $C^f(y_{k+1})$. Here, we assume that the minimal circuit $C^f$ determines the value $f(y_{k+1})$. Otherwise, there are two circuits $C^1, C^2$ of minimal size such that $C^1(y_{k+1})\ne C^2(y_{k+1})$, and therefore any prediction is equally good. Say that the size of the minimal circuit $C^f$ is $s$. Then the task to predict the value $C^f(y_{k+1})$ can be formulated as the task to prove an $s$-size circuit lower bound of the form $$\forall\text{ circuit }C\text{ of size }s,\ \bigvee_{i=1,\dots,k} C(y_i)\ne f(y_i)\vee C(y_{k+1})\ne\epsilon$$ for $\epsilon=0$ or $\epsilon=1$. 
\vspace{3pt}}}
\bigskip

An interesting aspect of the prediction method described in box A is that by proving even a single circuit lower bound we can learn something about the function $f$ (if we know the value $s$). More precisely, we predict $C^f$ on a single input but do not necessarilly gain knowledge of the values of $C^f$ on other inputs. This `instance-specific' learning should be contrasted with PAC learning, Definition \ref{d:lear}, where one is required to generate a circuit predicting the target function $f$ on most inputs. This, however, does not mean that it is easier to learn in the sense of box A: in Definition \ref{d:lear} we do not need to recognize when the prediction errs while the prediction from box A is zero-error in the sense that it guarantees to output the right value of $C^f(y_{k+1})$.\footnote{{\bf Provability vs truth.} The definition of `the right' prediction in terms of minimal circuits used in box A can be interpreted as an implicit (alternative) definition of truth. Consider, for example, that strings $y_j$ encode statements in set theory ZFC and the value $f(y_j)$ is 1 if and only if the statement encoded by $y$ is provable in ZFC.  It would be interesting to find out whether the minimal circuit coinciding with a sufficiently rich list of such samples $(y_j,f(y_j))$ determines a truth value of the Continuum Hypothesis or of the consistency of ZFC, statements which are independent of ZFC. Unfortunately, in general, such questions seem to be out of reach of the contemporary mathematics.}


\medskip

\noindent {\bf Determining minimal circuit size.} A drawback of the observation in box A is that it requires knowledge of the size $s$ of the minimal circuit $C^f$, which might be hard for the learner to determine. The size $s$ could be determined by deciding $t$-size circuit lower bounds for $t\in [s]$. Perhaps a more practical way of addressing the issue is to take a sufficiently big approximate value $s'$ of $s$, choose a random $t\in [s']$ and prove $t$-size lower bounds (as in box A with $t$ instead of $s$). If $s'\le n^{O(1)}$, the probability that we have the right $t$ is $1/n^{O(1)}$. Then, by solving polynomially many $t$-size lower bounds (in order to predict $C^f(y)$ on polynomially many $y$'s), we can approximate the accuracy of our predictions. If the accuracy is not high, we can reapeat the process with a new random $t\in [s']$. 
The advantage of this method is that it does not rely on deciding correctly whether some particular $t$-size circuit lower bounds hold - we are actually allowed to err on some fraction of lower bounds. However, its predictions are no longer zero-error. A closely related argument is formalized in Section \ref{s:leargen}.
\bigskip

\noindent {\bf Proof complexity.} The prediction method from box A 
relies on proof complexity of circuit lower bounds, cf. \cite{Kpc}.\footnote{Notably, Razborov \cite{Rkdnf} established that weak proof systems such as Resolution operating with $k$-DNFs for small $k$ do not have polynomial-size proofs of any superpolynomial circuit lower bound whatsoever and he conjectured this holds under a hardness assumption even for stronger systems such as Frege. The issue is, however, delicate because proof systems like Extended Frege are already capable of formalizing a lot of complexity theory, see e.g. \cite{MP}, and it is perfectly plausible that if a circuit lower bound is provable at all, then it is efficiently provable in Extended Frege.} It would be interesting to find out if proving circuit lower bounds in standard proof systems suffices to construct learning circuits.

\begin{question}[Learning interpolation]\label{q:fip}
Is there a p-time function which given an Extended Frege proof of a formula $\bigvee_{y\in A} C(y)\ne f(y)\vee C(x)\ne\epsilon$, for $\epsilon=0$ or $\epsilon=1$, with free variables representing $s$-size circuits $C$ with $n$ inputs, a fixed set $A$ of $n$-bit inputs of a sufficiently big size $|A|=poly(s,n)$, a fixed $n$-bit string $x\notin A$ and values of $f\in\Circuit[s]$ on $A$, outputs a circuit $(1/2+1/n)$-approximating $f$?
\end{question}

\subsection{Learning from witnessing lower bounds}\label{s:witnessing}

We now give a construction of PAC learning algorithms from an interactive witnessing of circuit lower bounds. As discussed in the introduction, the implication can be also interpreted as a construction of PAC learning algorithms from a frequent interactive instance-specific learning.

\medskip


\begin{theorem}[Learning from interactive witnessing of lower bounds]\label{t:main}
Let $d\ge 2; k,K \ge 1$ and $H$ be a Boolean function with $n$ inputs. 
Assume there are $2^{Kn}$-size circuits $W_1^1,\dots,W_{\log n}^b$ with $b=2^{Kn}$ such that for each distribution $\mathcal{R}$ on $n^{10dk}$-size circuits with $n$ inputs 
there exists $j\in [b]$ such that circuits $W_1^j,\dots,W_{\log n}^j$ witness errors of $n^{10dk}$-size circuits attempting to compute $H$ in the following way.
\begin{itemize} \item[] Given an oracle access to a random $n^{10dk}$-size circuit $D(x)$ with $n$ inputs, with probability at least $1-3/n^3$ over $\mathcal{R}$, the following interactive protocol succeeds: After querying values of 
circuit $D$, $W_1^j$ outputs a not-yet-queried $x_1\in\{0,1\}^n$ s.t. $D(x_1)\ne H(x_1)$ or $W_2^j$ receives a correction in the form of bits $D(x_1), H(x_1)$ s.t. $D(x_1)=H(x_1)$. Having $D(x_1), H(x_1)$ and the samples queried by $W^j_1$, $W_2^j$ makes further queries to $D$ and generates the second not-yet-queried candidate $x_2\in\{0,1\}^n$ for the claim $C(x_2)\ne H(x_2)$. If $D(x_2)=H(x_2)$, $W_3^j$ receives a correction and the protocol continues in this way until some $W_t^j$, for $t\le \log n$, with access to all previous corrections and samples finds the right $x_t$ which has not been queried by $W^j_1,\dots,W^j_t$ and witnesses $D(x_t)\ne H(x_t)$. 

\end{itemize}

Then, circuits of size $n^{dk}$ with $n^d$ inputs can be learned by circuits of size $2^{K'n}$ over the uniform distribution with non-adaptive membership queries, confidence $1/2^{K'n^{2}}$ up to error $1/2-1/2^{K'n^{2}}$, where $K'$ is a constant depending only on $K$.
\end{theorem}

Note that the witnessing circuits from Theorem \ref{t:main} can work for arbitrary function $H$ and, for the circuits $D$ on which the witnessing succeeds, the number of queries in each round is implicitly bounded by $<2^n$ (since after querying $D$ on all inputs it would be impossible to output a not-yet-queried input).


\proof The proof follows the main construction from \cite{Pclba, Knw} in the context of learning. The main technical complication is caused by the fact that the witnessing circuits $W^1_1\dots,W^b_{\log n}$ are allowed to fail on a significant fraction of inputs. 
\medskip




In order to derive the conclusion of the theorem it suffices to assume that the witnessing circuits work for distributions $\mathcal{R}$ induced by specific Nisan-Wigderson generators. 

Consider a Nisan-Wigderson generator based on a circuit $C$ which we aim to learn. Specifically, for $d\ge 2$ and $n^{2d}\le m\le 2n^{2d}$, let $A=\{a_{i,j}\}^{i\in [2^n]}_{j\in [m]}$ be a $2^n\times m$ 0-1 matrix with $n^{d}$ ones per row and $J_i(A):=\{j\in  [m]; a_{i,j}=1\}$. Then define an NW-generator $NW_{C}:\{0,1\}^{m}\mapsto\{0,1\}^{2^n}$ as $$(NW_{C}(w))_i=C(w|J_i(A))$$ where $w|J_i(A)$ are $w_j$'s such that $j\in J_i(A)$.

For any $d\geq 2$, Nisan and Wigderson \cite{NW} constructed a $2^n\times m$ 0-1 matrix $A$ with $n^{d}$ ones per row and $n^{2d}\le m\le 2n^{2d}$ which is also an $(n,n^{d})$-design meaning that for each $i\neq j$, $|J_i(A)\cap J_j(A)|\leq n$ and $|J_i(A)|=n^d$. Moreover, there are $n^{9d}$-size circuits which given $i\in \{0,1\}^n$ and $w\in\{0,1\}^{m}$ output $w|J_i(A)$, cf. \cite{CIKK}. Therefore, if $C$ has $n^{d}$ inputs and size $n^{dk}$, then for each $w\in \{0,1\}^{m}$, $(NW_{C}(w))_x$ is a function on $n$ inputs $x$ computable by circuits of size $n^{10dk}$. We want to learn $C$ by a circuit of size $2^{O(n)}$.

Let $\mathcal{R}$ be the distribution on $n^{10dk}$-size circuits defined so that a random circuit over $\mathcal{R}$ is $(NW_C(w))_x$ for $w\in\{0,1\}^m$ chosen uniformly at random.
\medskip

By the assumption of the theorem, we have $2^{Kn}$-size circuits $W_1^1,\dots, W_{\log n}^b$, with $b=2^{Kn}$ such that for some $j\in [b]$ for $1-3/n^3$ of all $w\in\{0,1\}^{m}$ 
circuits $W_1^j,\dots, W_{\log n}^j$ find an error of the $n^{10dk}$-size circuit $(NW_C(w))_x$ attempting to compute $H$. We will use them in order to break, in a certain sense, the generator $NW_C$ and reconstruct the circuit $C$. 

For each $w$ define a trace $tr(C,w)=x_1,\dots, x_t$ as the sequence of $t\le \log n$ strings generated by $W_1^j,\dots,W_t^j$ on $(NW_C(w))_x$ such that $W_t^j$ is the first circuit which succeeds in witnessing the error, i.e. $H(x_t)\ne (NW_C(w))_{x_t}$. If circuits $W^j_1,\dots,W^j_{\log n}$ do not find an error, $x_t=x_{\log n}$. The trace is defined w.r.t. a fixed `helpful' oracle $Y$ providing corrections in the form of bits $(NW_C(w))_x, H(x)$. 

\medskip
For $u\in \{0,1\}^{n^{d}}$ and $v\in\{0,1\}^{m-n^{d}}$ define $r_x(u,v)\in \{0,1\}^{m}$ by putting bits of $u$ into positions $J_x(A)$ and filling the remaining bits by $v$ (in the natural order). We say that $w\in \{0,1\}^{m}$ is {\em good} if the trace $tr(C,w)$ ends with a string witnessing an error of circuit $(NW_C(w))_x$ and {\em bad} otherwise. Similarly, given $v\in \{0,1\}^{m-n^d}$ and $x'\in \{0,1\}^n$, we say that $u\in\{0,1\}^{n^d}$ is good if $r_{x'}(u,v)$ is.

The core claim of the proof is the existence of a frequent trace on which circuit $W^j_1,\dots,W^j_{\log n}$ succeed in witnessing the error with significant advantage.

\begin{claim}\label{claim} 
There is a trace $Tr=X_1,\dots, X_t, t\leq \log n$ such that for $s\ge 1/(6^{2n(t-1)}2^{2n}n)$ of all $a\in \{0,1\}^{m-n^{d}}$ for $s'\ge s$ of all $u\in\{0,1\}^{n^d}$ $tr(C,r_{X_t}(u,a))$ starts with $Tr$ and at least $(2/3-6^t/n^3-2/n)s'2^{n^d}$ $u$'s are good and satisfy $tr(C,r_{X_t}(u,a))=Tr$.
\end{claim}

The trace $Tr$ is constructed inductively: in step $i$ we want to find $X_1,\dots, X_{i-1}$ such that for $\ge 1/6^{2n(i-1)}$ of all $w$'s $tr(C,w)$ strictly extends $X_1,\dots,X_{i-1}$ and the fraction of good $w$'s for which this happens is $\ge 1-6^i/2n^3$. For $i=1$ this holds by the assumption. Assume we have such $X_1,\dots, X_{i-1}$. We want to extend them to $X_1,\dots,X_i$. Since there are at most $2^n$ strings $X_j$, there is $X_i$ such that for $s''\ge 1/(2^{2n}6^{2n(i-1)})$ $w$'s $tr(C,w)$ starts with $X_1,\dots,X_i$ and $\le 6^i/n^3$ of these $w$'s are bad. Otherwise, the fraction of good $w$'s for which $tr(C,w)$ strictly extends $X_1,\dots,X_{i-1}$ would be $\le 1/2^n+1-6^i/n^3<1-6^i/2n^3$ if $2n^3\le 2^n$. Now, either for $\ge (2/3)s''$ of $w$'s $tr(C,w)$ stops at $X_i$ (hence, for $\le (1/3)s''$ $w$'s the trace continues and for $\le 6^is''/n^3$ bad $w$'s $tr(C,w)$ starts with $X_1,\dots,X_i$) or for $\ge (1/3)s''$ $w$'s the trace strictly extends $X_1,\dots,X_i$. In the latter case, for $\le 6^is''/n^3$ bad $w$'s $tr(C,w)$ starts with $X_1,\dots,X_i$, which means that the fraction of bad $w$'s such that $tr(C,w)$ strictly extends $X_1,\dots,X_i$ is $\le 3\cdot 6^i/n^3$.

Since for all $w$, the length of $tr(C,w)$ is bounded by $\log n$, the process of extending $X_1,\dots, X_{i-1}$ has to stop at some step $1\le i\le \log n$. That is, there is $Tr=X_1,\dots, X_t, t\le \log n$ such that for $\ge (2/3)s$ of $w$'s $tr(C,w)=Tr$, for $\le (1/3)s$ of $w$'s $tr(C,w)$ strictly extends $Tr$ and $\le 6^ts/n^3$ of $w$'s such that $tr(C,w)$ is consistent with $Tr$ are bad, where $s\ge 1/(6^{2n(t-1)}2^{2n})$. 
The number of good $w$'s such that $tr(C,w)=Tr$ is at least $(2/3-6^t/n^3)s2^m$. Therefore, $\ge s/n$ $a$'s can be completed by $s'\ge s/n$ $u$'s to a string $w=r_{X_t}(u,a)$ such that $tr(C,w)$ starts with $Tr$ and at least $(2/3-6^t/n^3-2/n)s'2^{n^d}$ $u$'s are good and satisfy $tr(C,r_{X_t}(u,a))=Tr$. This proves the claim.
\medskip

For $X\in \{0,1\}^n$ and $a'\in\{0,1\}^{m-n^d}$ let $r_{X}(\cdot,a')$ be the bits of $a'$ in the positions of $[m]\backslash J_{X}(A)$. 
Since $A$ is an $(n,n^{d})$-design, for any row $x\neq X$ at most $n$ bits of $r_{X}(\cdot,a')|J_x(A)$ are not set. 
For 
$x\ne X$, let $Y_{x,C}^{X,a'}$ be the set of all corrections provided by $Y$ on $x, C$ and $r_{X}(u,a')|J_x(A)$ for all $u\in\{0,1\}^{n^{d}}$. This includes queries to $C$ on inputs $r_{X}(u,a')|J_x(A)$. The size of each set $Y_{x,C}^{X,a'}$ is $2^{O(n)}$. 
\smallskip

We are ready to describe a circuit $D'$ that approximates $C$. First, choose uniformly at random $a'\in\{0,1\}^{m-n^d}$, a trace $X^1,\dots,X^t$ with $t\le \log n$, a bit $maj\in \{0,1\}$ and $j'\in [b]$. Query $C$ so that all queries to $C$ from sets $Y_{x,C}^{X^t,a'}$, for $x\ne X^t$, are obtained. In order to get access to all corrections from $Y_{X^1,C}^{X^t,a'},\dots, Y_{X^{t-1},C}^{X^t,a'}$ we provide also the full truth-table of $H$ as a nonuniform advice of $D'$. The truth table of $H$ is a single nonuniform advice of the learner which works for every $C$. Then $D'$ computes as follows.
For each $u\in\{0,1\}^{n^{d}}$ produce $r_{X^t}(u,a')$. 
Next, use $W_1^{j'}$ to produce $x^1$. If a query of $W_1^{j'}$ cannot be answered by $Y_{x,C}^{X^t,a'}$ with $x\ne X^t$ or  $x^1\ne X^1$, output $maj$. Otherwise, use the advice from $Y_{X^1,C}^{X^t,a'}$ to find out if $H(X^1)=NW_{C}(r_{X^t}(u,a'))_{X^1}$. If the equality does not hold, output $maj$. Otherwise, use $W_2^{j'}$ to generate $x^2$ and continue in the same manner until $W_t^{j'}$ produces $x^t$. If a query of $W_t^{j'}$ cannot be answered by $Y_{x,C}^{X^t,a'}$ with $x\ne X^t$ or $x^t\ne X^t$, output $maj$. Otherwise, output 0 iff $H(X^t)=1$. The resulting circuit $D'$ has $n^{d}$ inputs and size $2^{O(n)}$, if $m\le 2^n$ (which holds w.l.o.g.).
\medskip

By Claim \ref{claim}, with probability at least $1/(6^{2n\log n}2^{O(n\log n)})$ the learner guessed $j'=j$, trace $Tr$ and assignment $a$ such that for at least $(2/3-6^t/n^3-2/n)s'$ of all $u\in\{0,1\}^{n^{d}}$, $D'$ will successfully predict $C(u)$. Moreover, for at most $(1/3+6^t/n^3+2/n)s'$ of all $u$'s, the trace extends $Tr$ or starts with $Tr$ but does not end with a string witnessing an error. Since with probability $1/2$ the correct value on at least half of all remaining $u$'s is $maj$, $\Pr_u[D'(u)=C(u)]\geq 1/2+(1/6-6^t/n^3-2/n)s$.
 \qed

\bigskip

The assumption from Theorem \ref{t:main} is justified by the following lemma which establishes the converse.

\begin{lemma}[Witnessing from learning]\label{l:mainconverse}
Let $k\ge 1$; $\epsilon<1$; $2^n/2n\ge 2^{\epsilon n}\ge n^k$ and $H$ be a Boolean function with $n$ inputs hard to $(1-1/n)$-approximate by circuits of size $2^{\epsilon n}$. 
Assume $\Circuit[n^k]$ can be learned by $\Circuit[2^{\epsilon n}]$ over the uniform distribution with confidence $1$ up to error $\epsilon'$.

Then, there are $2^{O(n)}$-size circuits $W^1,\dots,W^b$ with $b=2^n/2n$ such that for each distribution $\mathcal{R}$ on $n^{k}$-size circuits with $n$ inputs 
there exists $j\in [b]$ such that given an oracle access to a random $n^{k}$-size circuit $D(x)$ with $n$ inputs, with probability at least $1-2\epsilon' n$ over $\mathcal{R}$, after $\le 2^{\epsilon n}$ queries to 
circuit $D$, $W^j$ outputs a not-yet-queried $x\in\{0,1\}^n$ s.t. $D(x)\ne H(x)$.

\end{lemma}

\proof By the assumption, there exists an $2^{\epsilon n}$-size circuit $W$ which for each $n^k$-size circuit $D$, given an oracle access to $D$, outputs a circuit $C$ $(1-\epsilon')$-approximating $D$. Since $H$ is hard to $(1-1/n)$-approximate by circuits of size $2^{\epsilon n}\le 2^n/2n$, there are at least $2^n/2n$ inputs which have not been queried by $W$ and on which $C$ fails to compute $H$. Therefore, a random input which has not been queried by $W$ and on which $C$ fails to compute $H$ witnesses $D(x)\ne H(x)$ with probability $\ge 1-2\epsilon' n$. Let $W^1,\dots,W^b$, $b=2^n/2n$, be circuits such that $W^i$ simulates $W$ and outputs the $i$-th input on which $C$ fails to compute $H$ ignoring inputs which have been queried by $W$. The size of each $W^i$ is $2^{O(n)}$ because it uses the whole truth table of $H$ as a nonuniform advice. Let $\mathcal{R}$ be arbitrary distribution on circuits of size $n^k$. Since for each $D$, at least $1-2\epsilon' n$ of $W^i$'s succeed, there is $W^j$ which succeeds on random $D$ with probability $\ge 1-2\epsilon' n$ over $\mathcal{R}$.
\qed

\bigskip

Note that Theorem \ref{t:main} together with Lemma \ref{l:mainconverse} imply that for suitable $H$ it is possible to collapse the number of rounds in the interactive witnessing from Theorem \ref{t:main} at the expense of witnessing errors of slightly smaller circuits (and a small increase in the running time of the witnessing).

\def\backup{

\begin{theorem}[Instance-specific learning from interactive witnessing of lower bounds]\label{t:main}
Let $d\ge 2, k\ge 1$ and $H$ be an \NP machine. Assume there are $poly(n)$-size circuits $W_1,\dots,W_{l}$ for $l=O(n)$ which witness errors of $n^{5dk}$-size circuits attempting to compute $H$ in the following way: After $poly(n)$ queries to an $n^{5dk}$-size circuit $D$, $W_1$ outputs $x_1\in\{0,1\}^n$ such that $D(x_1)\ne H(x_1)$ or $W_2$ receives a counterexaple to its claim in the form of $D(x_1)$ and $y_1\in\{0,1\}^n$ s.t. $D(x_1)=1$ iff $H(x_1)$ accepts $x_1$ with nondeterministic bits $y_1$. Having $D(x_1)$ and $y_1$, $W_2$ generates the second candidate $x_2\in\{0,1\}^n$ for the claim that $C(x_2)\ne H(x_2)$. If $D(x_2)=H(x_2)$, $W_3$ receives a counterexample and the protocol continues in this way until some $W_t$, for $t\le l$, with access to all previous counterexamples finds the right $x_l$ witnessing $D(x_l)\ne H(x_l)$. 

Then, $\Circuit[n^{k}]$ can be learned by $\Circuit[2^{O(n^{1/d})}]$ with confidence 1 and error $1/2-1/2^{O(n^{2/d})}$.
\end{theorem}

\proof The proof adapts the main construction from \cite{Pclba, Knw} to the context of learning. We describe $2^{O(n)}$-size circuits learning circuits of size $n^{dk}$ with $n^{d}$ inputs, for $d\ge 2$. 
\smallskip

Consider a Nisan-Wigderson generator based on a circuit $C$ which we aim to learn. Specifically, for $d\ge 2$ and $n^{2d}\le m\le 2n^{2d}$, let $A=\{a_{i,j}\}^{i\in [2^n]}_{j\in [m]}$ be $2^n\times m$ 0-1 matrix with $n^{d}$ ones per row and $J_i(A):=\{j\in  [m]; a_{i,j}=1\}$. 
Then define an NW-generator $NW_{C}:\{0,1\}^{m}\mapsto\{0,1\}^{2^n}$ as $$(NW_{C}(w))_i=C(w|J_i(A))$$ where $w|J_i(A)$ are $w_j$'s such that $j\in J_i(A)$.
\smallskip

For any $d\geq 2$, Nisan and Wigderson \cite{NW} constructed $2^n\times m$ 0-1 matrix $A$ with $n^{d}$ ones per row and $n^{2d}\le m\le 2n^{2d}$ which is also an $(n,n^{d})$-design meaning that for each $i\neq j$, $|J_i(A)\cap J_j(A)|\leq n$. Moreover, there are $n^{4d}$-size circuits which given $i\in \{0,1\}^n$ and $w\in\{0,1\}^{m}$ output $w|J_i(A)$, cf. \cite{CIKK}. Therefore, if $C$ has $n^{d}$ inputs and size $n^{dk}$, then for each $w\in \{0,1\}^{m}$, $(NW_{C}(w))_x$ is a function on $n$ inputs $x$ computable by circuits of size $n^{5dk}$.
\medskip

By the assumption of the theorem, there are $l=O(n)$ circuits $W_1,\dots, W_l$ of $poly(n)$-size which for each $w\in\{0,1\}^{m}$ find an error of the $n^{5dk}$-size circuit $(NW_C(w))_x$ attempting to compute $H$. Define a trace $tr(w)=x_1,\dots, x_t$ as the sequence of $t\le l$ strings generated by $W_1,\dots,W_t$ such that $W_t$ is the first circuit which succeeds in witnessing the error, i.e. $H(x_t)\ne (NW_C(w))_{x_t}$. The trace is defined w.r.t. a fixed `helpful' oracle $Y$ providing counterexamples in the form of bits $(NW_C(w))_x$ and strings $y$ such that $H$ accepts $x$ with nondeterministic bits $y$, if $H$ accepts at all. The advice provided by $Y$ depends only on $x$ and $w|J_x(A)$.

\medskip
For $u\in \{0,1\}^{n^{d}}$ and $v\in\{0,1\}^{m-n^{d}}$ define $r_x(u,v)\in \{0,1\}^{m}$ by putting bits of $u$ into positions $J_x(A)$ and filling the remaining bits by $v$ (in the natural order). 

\begin{claim} There is a trace $Tr=X_1,\dots, X_t, t\leq l$ and $a\in \{0,1\}^{m-n^{d}}$ such that $Tr=tr(r_{x_t}(u,a))$  for at least $2/(3(2^n))^t$ of all $u$'s. 
\end{claim}

$Tr$ and $a$ are constructed inductively: Since there are at most $2^n$ strings $X_j$, there is $X_1$ such that for at least $1/2^n$ of all $w$'s $tr(w)$ begins with $X_1$. Either there is $a\in \{0,1\}^{m-n^{d}}$ such that $tr(r_{X_1}(u,a))=X_1$ for at least $2/(3(2^n))$ of all $u$'s or there is $X_2$ such that for at least $1/(3(2^{2n}))$ of all $w$'s $tr(w)$ begins with $X_1,X_2$. Assume that for at least $1/(3^{i-1}(2^{in}))$ of $w$'s $tr(w)$ begins with $X_1,\dots, X_i$. Either there is $a\in \{0,1\}^{m-n^{d}}$ such that $tr(r_{X_{i}}(u,a))=X_1,\dots,X_i$ for at least $2/(3(2^n))^i$ of all $u$'s or there is $X_{i+1}$ such that for at least $1/(3^i(2^{(i+1)n}))$ $w$'s $tr(w)$ begins with $X_1,\dots,X_{i+1}$. This proves the claim.
\medskip

Fix now $Tr, a$ from the claim and let $r_{X_t}(\cdot,a)$ be the bits of $a$ in the positions of $[m]\backslash J_{X_t}(a)$. 
\smallskip

Let $u\in \{0,1\}^{n^{d}}$. Since $A$ is $(n,n^{d})$-design, for any row $x\neq X_t$ at most $n$ bits of $r_{X_t}(\cdot,a)|J_x(A)$ are not set. 
Let $Y_x, x\neq X_t$ be the set of all counterexamples provided by $Y$ on $x$ and $r_{X_t}(u,a)|J_x(A)$ for all $u\in\{0,1\}^{n^{d}}$. This includes queries to $C$ on inputs $r_{X_t}(u,a)|J_x(A)$. The size of each set $Y_x$ is $2^{O(n)}$. 
\smallskip

We are ready to describe a circuit $D'$ that approximates $C$ using as advice $X_1,\dots,X_t$, $r_{X_t}(\cdot,a)$ and sets $Y_{X_1},\dots,Y_{X_{t-1}}$. 
For each $u\in\{0,1\}^{n^{d}}$ produce $r_{X_t}(u,a)$. 
Then use $W_1$ to produce $x^1$. If $x^1\ne X_1$, output a random bit. Otherwise, use the advice from $Y_{X_1}$ to find out if $H(X_1)=NW_{C}(r_{X_t}(u,a))_{X_1}$. If the equality does not hold, output a random bit. Otherwise, use $W_2$ to generate $x^2$ and continue in the same manner until $W_t$ produces $x^t$. If $x^t\ne X_t$, output a random bit. Otherwise, output 0 iff $H(X_t)=1$. The resulting circuit $D'$ has $n^{d}$ inputs and size $2^{O(n)}$.
\medskip

By the choice of $Tr$, for at least $2/(3(2^n))^t$ of all $u\in\{0,1\}^{n^{d}}$, $D'$ will successfully predict $C(u)$. 
Moreover, at most $1/(3(2^n))^t$ of all traces extend $X_1,\dots,X_t$. Since a random bit is the correct value on at least half of all $u$'s for which $tr(r_{X_t}(u,a))$ is not $Tr$ and does not extend $Tr$, $\Pr_u[D'(u)=C(u)]\geq 1/2+1/(3^{t}2^{nt+1})$. 
 \qed

}

\def\lfromisl{
A direct consequence of Theorem \ref{t:main} is that if we are given a learning algorithm which is able to predict a value of a significant fraction of p-size circuits w.r.t. a suitable distribution after polynomially many queries even on a single input, this already implies learnability of p-size circuits on almost all inputs. Moreover, if the prediction is done in a reliable way as described in box A, i.e. the prediction is zero-error, the learner is zero-error on a significant fraction of inputs with a significant confidence.

\begin{corollary}[Learning from frequent instance-specific learning]\label{c:main}
Let $d\ge 2, k\ge 1$ and $H$ be Boolean function on $n$ inputs. Assume there are $2^{O(n)}$-size circuits $W_1^1,\dots,W_{l}^b$ for $l=O(\log n), b=2^{O(n)}$ witnessing errors of $n^{10kd}$-size circuits $(NW_C(w))_x$ as in Theorem \ref{t:main}. Further, assume that each circuit from $W_1^1,\dots,W_l^b$ can recognize before receiving corrections if its output is going to work.\textcolor{red}{This is like having no interactions.}

Then, there are $2^{O(n)}$-size circuits which make $poly(n)$ queries to the $n^{dk}$-size circuit $C$ with $n^d$ inputs and predict the value of $C$ with zero-error on a $1/2^{O(n^{5})}$ fraction of inputs with confidence $1/2^{O(n^{5})}$.
\end{corollary}}

\medskip

\noindent {\bf Learning from witnessing lower bounds with white-box access.} Theorem \ref{t:main} holds also under the stronger assumption that circuits $W^1_1\dots, W^b_{\log n}$ witness errors of $n^{10dk}$-size nondeterministic circuits $D$ with $n$ inputs (and $\le n^{10dk}$ nondeterministic bits), where $D$ computes a function in $\Circuit[n^{10dk}]$, i.e. $D$ is a nondeterministic circuit computing a function in \Ppoly. Then it makes sense to allow $W^1_1,\dots, W^b_{\log n}$ to access a full description of a given nondeterministic circuit $D$. The conclusion of the resulting theorem remains valid with the only difference that the learning algorithm is given full description of an $n^{dk}$-size nondeterministic circuit with $n^d$ inputs representing the target function (which is computable by an $n^{dk}$-size deterministic circuit with $n^d$ inputs). 
\medskip

\noindent {\bf Comparison to witnessing in bounded arithmetic.} The existence of witnessing analogous to the one from Theorem \ref{t:main} follows from the provability of circuit lower bounds in bounded arithmetic.

If $H:\{0,1\}^n\rightarrow \{0,1\}$ is an \NP function and $n_0, k$ are constants, we can write down a $\forall\Sigma^b_2$ formula $\LB(H,n^k)$ stating that $H$ is hard for circuits of size $n^k$: $$\forall n,\ n>n_0\ \forall\ \text{circuit}\ D\ \text{of size}\ \leq n^k\ \exists y,\ |y|=n,\ D(y)\neq H(y),$$ where $D(y)\neq H(y)$ is a $\Sigma^b_2$ formula stating that a circuit $D$ on input $y$ outputs the opposite value of $H(y)$. Here, $\Sigma^b_2$ is a class of formulas in the language of Cook's theory \PV which define precisely the predicates from $\Sigma^p_2$ level of the polynomial hierarchy, cf.~\cite{Kpc}. 

By the KPT theorem \cite{KPT}, if \PV proves $\LB(H,n^k)$ then there are finitely many $poly(n)$-time functions $W_1,\dots, W_l$ which witness the existential quantifiers of $\LB(H,n^k)$ (including the existential quantifier from the subformula $D(y)\ne H(y)$) in the same interactive way as in Theorem \ref{t:main} except that the corrections include strings standing for the innermost universal quantifier of $\LB(H,n^k)$ (which allow to verify in p-time that $D(y)\ne H(y)$ has not been witnessed by the most recent candidates). Moreover, $W_1,\dots, W_l$ have access to the full description of a given circuit $D$ and do not make queries to $D$ but directly generate potential errors, cf. \cite{Pclba}. 
 
It is possible to change the formula $\LB(H,n^k)$ by introducing a parameter $m$ satisfying $2^n=|m|$ so that the witnessing from the \PV-provability of the new formula is given by circuits $W_1,\dots, W_l$ of size $2^{O(n)}$. In such case, $H$ is allowed to be in $\mathsf{NE}$. We could allow $H$ to be even an arbitrary Boolean function if we formulated the lower bound in QBF proof systems instead of bounded arithmetic. 

A crucial difference between the black-box witnessing from Theorem \ref{t:main} and white-box witnessing in bounded arithmetic is that, under standard hardness assumptions, the white-box witnessing of p-size circuit lower bounds for functions $H$ such as \SAT exists, cf. \cite{MP}. 


\medskip
\noindent {\bf Comparison to other witnessing theorems.} Lipton and Young \cite{LY} showed that for each Boolean function $H$ hard for circuits of size $O(n^{k+1})$ there is a multiset of inputs $A$ of size $O(n^k)$, the so called anticheckers, such that each $n^k$-size circuit fails to compute $H$ on $\ge 1/3$ of inputs from $A$.
Therefore, for each distribution $\mathcal{R}$ on $n^k$-size circuits, some input from the set of anticheckers will witness an error of a random $n^k$-size circuits $D$ (without a single query to $D$) with probability $\ge 1/3$ over $\mathcal{R}$. Using $t$ rounds the probability of witnessing an error can be increased to $1-1/(3/2)^t$. This can be done with $\le n^{O(kt)}$ witnessing circuits $W^i_j$. More precisely, we can let $W^i_1,\dots,W^i_t$ to be the $i$-th possible $t$-tuple of inputs from the set of anticheckers, for $i<n^{O(kt)}$. Theorem \ref{t:main} shows that it is not possible to increase this probability further to $1-3/n^3$ using $\log n$ rounds unless p-size circuits can be learned efficiently. 

Gutfreund, Shaltiel and Ta-Shma \cite{GST} showed that if \Ptime $\ne$ \NP there is a p-time algorithm which, given a description of an $n^k$-time machine $D$, generates a set of $\le 3$ formulas such that $D$ fails to solve \SAT on one of them. Atserias \cite{Abb} extended this by showing that if $\NP\not\subseteq \mathsf{BPP}$ there is a probabilistic p-time algorithm which, given an oracle access to an $n^k$-time machine $D$, outputs with probability $\ge 1/8$ a set of formulas such that $D$ fails to solve \SAT on one of them. These algorithms differ from the witnessing in Theorem \ref{t:main} in several ways: they find errors of uniform algorithms, are allowed to generate errors of different lengths, generate errors with a significantly smaller probability than the probability required in Theorem \ref{t:main} and the set of formulas generated by the algorithm of Atserias includes formulas on which the algorithm queried $D$.

\section{Learning from breaking pseudorandom generators}\label{s:leargen}

Circuit lower bounds can be used to construct PAC learning algorithms also if we assume that they break pseudorandom generators. The construction goes back to a 
relation between predictability and pseudorandomness which can be interpreted in terms of learning algorithms, as shown by Blum, Furst, Kearn and Lipton \cite{BFKL} and later extended by several other works. In this section we survey some of these connections, derive a construction of learning algorithms from the non-existence of succinct nonuniform pseudorandom function families and show how these connections relate to a question of Rudich about turning demibits to superbits.
\bigskip

We start by recalling the construction from \cite{BFKL}, which underlies all results in this section.

\medskip
For an $n^c$-size circuit $C$ with $n$ inputs define a generator $$G_C:\{0,1\}^{mn}\mapsto\{0,1\}^{mn+m}$$ which maps $m$ $n$-bit strings $x_1,\dots,x_m$ to $x_1,C(x_1),\dots,x_m,C(x_m)$. 

\begin{lemma}[from \cite{BFKL}]\label{l:gen}
There is a randomized p-time function $L$ such that for every $n^c$-size circuit $C$, 
if an $s$-size circuit $D$ satisfies $$\Pr[D(x)=1]-\Pr[D(G_C(x))=1]\ge 1/s,$$ then the circuit $C$ is learnable by $L(D)$ over the uniform distribution with random examples, confidence $1/2m^2s$, up to error $1/2-1/2ms$.
\end{lemma}

\proof Given $D$, $L(D)$ chooses a random $i\in [m]$, random bits $r_{i},\dots,r_m$, random $n$-bit strings $x_1,\dots,x_n$ except $x_i$ and queries the bits $C(x_1),\dots, C(x_{i-1})$. For $x_i\in\{0,1\}^n$, let $p_i:=D(x_1,C(x_1),\dots,x_{i-1},C(x_{i-1}),x_i,r_i,\dots,x_m,r_m)$. Then $L(D)$ on $x_i$ predicts the value $C(x_i)$ by outputting $\neg r_i$ if $p_i=1$ and $r_i$ otherwise. By triangle inequality, random $i\in [m]$ satisfies $$\Pr[p_{i}=1]-\Pr[p_{i+1}=1]\ge 1/ms$$ with probability $1/m$. Since the probability over $r_i\dots,r_m,x_1,\dots,x_m$ that $L(D)$ predicts $C(x_i)$ correctly is $$\frac{1}{2}\Pr[p_i=1\mid r_i\ne C(x_i)]+\frac{1}{2}(1-\Pr[p_i=1\mid r_i= C(x_i)]),$$ and $\Pr[p_i=1]=\frac{1}{2} \Pr[p_i=1\mid r_i=C(x_i)]+\frac{1}{2}\Pr[p_i=1\mid r_i\ne C(x_i)],$ it follows that $$\Pr_{x_i}[L(D)(x_{i})=C(x_i)]\ge 1/2+1/2ms$$ with probability $1/2m^2s$ over the internal randomness of $L(D)$. \qed
\medskip


\bigskip

The proof of Lemma \ref{l:gen} implies that learning on average follows from breaking pseudorandom generators. Specifically, let $R$ be a p-size circuit which given $r$ bits outputs an $n^c$-size circuit $C$ and consider a generator $G:\{0,1\}^{mn+r}\mapsto \{0,1\}^{mn+m}$ which applies $R$ on its first $r$ input bits in order to output a circuit $C$ and then computes as a generator $G_C$ on the remaining $mn$ inputs. Breaking $G$ implies that we can break $G_C$ with significant probability over $C$ drawn from the distribution induced by $R$. Consequently, breaking $G$ means that we can learn a big fraction of $n^c$-size circuits w.r.t. $R$. Can we improve this average-case learning into a worst-case learning which works for all $n^c$-size circuits? Since efficient learning algorithms for p-size circuits yield natural properties useful against p-size circuits, which by \cite{RR} break pseudorandom generators, a positive answer would present an important dichotomy: cryptographic pseudorandom generators do not exist if and only if there are efficient learning algorithms for small circuits (with suitable parameters). This possibility has been explored by Oliveira-Santhanam \cite{OS} and Santhanam \cite{S19}, cf. Section \ref{s:prfs}.

\begin{question}[Dichotomy]\label{q:dichotomy} Assume that for each $\epsilon<1$ there is no pseudorandom generator $g:\{0,1\}^n\mapsto \{0,1\}^{n+1}$ computable in \Ppoly and safe against circuits of size $2^{n^{\epsilon}}$ for infinitely many $n$. Does it follow that p-size circuits are learnable by circuits of size $2^{O(n^\delta)}$, for some $\delta<1$, with confidence $1/n$, up to error $1/2-1/2^{O(n^\delta)}$?
\end{question}

\subsection{Worst-case learning from strong lower bound methods} 

The proof of Lemma \ref{l:gen} shows also that we can construct a worst-case learning algorithm assuming that given an oracle access to a pseudorandom generator we can efficiently produce its distinguisher. In particular, a single method breaking all pseudorandom generators would suffice.

\begin{definition} The circuit size problem $\GCSP[s,k]$ is the problem to decide whether for a given list of $k$ samples $(y_i,b_i)$, $y_i\in\{0,1\}^n, b_i\in\{0,1\}$, there exists a circuit $C$ of size $s$ computing the partial function defined by samples $(y_i,b_i)$, i.e. $C(y_i)=b_i$ for the given $k$ samples $(y_i,b_i)$. The parameterized minimum circuit size problem $\MCSP[s]$ stands for $\GCSP[s,2^n]$ where the list of $2^n$ samples defines the whole truth-table of a Boolean function.
\end{definition}

If we were extraordinary in proving circuit lower bounds, we could solve \GCSP efficiently. Note that $\MCSP[n^{O(1)}]\in\Ppoly$ is stronger assumption than the existence of \Ppoly-natural property useful against \Ppoly, which breaks pseudorandom generators.

The following theorem appeared (in different terminology) in Vadhan \cite{LR}, see also~\cite{ILO}.


\begin{theorem}[Learning from succinct natural proofs]\label{t:gcsplear}
Assume $\GCSP[n^c,n^{d}]\in\Ppoly$ for constants $d>c+1$. Then, $\Circuit[n^c]$ is learnable by \Ppoly over the uniform distribution with random examples, confidence $1/poly(n)$, up to error $1/2-1/poly(n)$.
\end{theorem}

\proof 
As the number of partial Boolean functions on a given set of $m$ inputs is $2^m$ and the number of $n^c$-size circuits is bouded by $2^{n^{c+1}}$, 
$\GCSP[n^c,n^{d}]\in\Ppoly$ implies that for $m=n^{d}$ there are p-size circuits $D$ such that for each $n^c$-size circuit $C$, $$\Pr[D(x)=1]-\Pr[D(G_C(x))=1]\ge 1/2.$$ Now, it suffices to apply Lemma \ref{l:gen}. \qed 

\def\susanna{
\bigskip

An instance-specific version of Theorem \ref{t:gcsplear} in which the prediction of $C(x)$ on a single input $x$ can be obtained from proving a single circuit lower bound holds by definition if we want to predict the value of $C(x)$ only if it is determined by the queried/given samples as in box A.\footnote{The proof of Theorem \ref{t:gcsplear} does not give us an instance-specific learning - it exploits the fact that suitable circuit lower bounds yield more often a correct prediction than an incorrect one.} We conclude the section by pointing out that this could be strengthened to an instance-specific construction of a learning algorithm which predicts $C(x)$ on many inputs assuming an error concentration hypthesis.

Suppose the size of circuit $C$ is $s=n^k$. A simple counting argument shows that if we chose randomly $y_1,\dots, y_{n^{k_1}}$ for a sufficiently big $k_1$, then with high probability each circuit of size $n^{k}$ which coincides with $C$ on $y_1,\dots y_{n^{k_1}}$ differs from $C$ only on a $\le \frac{1}{n}$-fraction of inputs. However, it is possible that each circuit of size $n^{k}$ errs on a different set of inputs and that their collective set of errors is too big: if the function computed by $C$ can be actually computed by a circuit of size $s-O(n)$, then for each $x$, we can construct a circuit $C'$ such that $C'(x)\ne C(x)$ but still $C'(y_i)=C(y_i)$ for $i=1,\dots,n^{k_1}$. Is this the case also if the size $s$ is guaranteed to be the size of a minimal circuit computing the partial function $y_1,C(y_1),\dots,y_{n^{k_1}},C(y_{n^{k_1}})$?

\begin{question}[Error concentration]\label{q:ach} Let $s:\mathbb{N}\mapsto\mathbb{N}$ be a function. Is it true that for every Boolean function $f$ with (minimal) circuit complexity $s$, for each $n$, there is $m=poly(s,n)$ such that randomly chosen $n$-bit strings $y_1,\dots,y_m$ force the set of $x\in\{0,1\}^n$ satisfying $C(x)\ne f(x)$ for some circuit $C$ of size $s$ which coincides with $f$ on $y_1,\dots,y_m$ to be small with high probability? More precisely, does the following hold? $$\Pr_{y_1,\dots,y_m}[|\{x\in\{0,1\}^n \mid \exists\ s\text{-size }C\text{ s.t.}\bigwedge_{y_1,\dots,y_m} C(y_i)=f(y_i)\wedge C(x)\ne f(x)\}|\le \frac{2^n}{n}]\ge 1-\frac{1}{n}$$
\end{question}

Since the size $s$ of the circuit $C$ can be provided as a nonuniform advice, a positive answer to Question \ref{q:ach} would mean that for every Boolean function $f$ whose minimal circuit has size $s$, for many inputs $x$, we can predict $f(x)$ by chosing random strings $y_1,\dots,y_m$ and proving a single lower bound as in box A. 
\textcolor{red}{To be erased: instance-specific part. Question 3 resolved by a simple counter-example of Erfan and Susanna. Unclear what would be the right formulation. Not so important anyway.}}

\subsection{Worst-case learning from natural proofs}\label{s:natcikk}

In Theorem \ref{t:gcsplear}, we can learn $f\in\Circuit[n^c]$ even if the algorithm for \GCSP works just for a significant fraction of partial truth-tables $(y_1,b_1),\dots,(y_{n^d},b_{n^d})$ with zero-error on easy partial truth-tables.  
Carmosino, Impagliazzo, Kabanets and Kolokolova \cite{CIKK} proved that the assumption of Theorem \ref{t:gcsplear} can be weakened to the existence of a standard natural property. The price for this is that the resulting learning uses membership queries instead of random examples. 
The crucial idea is similar to the proof of Theorem \ref{t:main}: apply the natural property (as an algorithm for suitable \GCSP) on a Nisan-Wigderson generator $NW_f$ based on the function $f$, which we want to learn.

\begin{theorem}[Learning from natural proofs \cite{CIKK}]\label{t:cikk}
Let $R$ be a $\Ppoly$-natural property useful against $\Circuit[n^d]$ for some $d \geq 1$. Then, for each $\gamma\in (0,1)$, $\Circuit[n^k]$ is learnable by $\Circuit[2^{O(n^{\gamma})}]$ over the uniform distribution with non-adaptive membership queries, confidence 1, up to error $\frac{1}{n^k}$, where $k = \frac{d\gamma}{a}$ and $a$ is an absolute constant.
\end{theorem}


\subsection{Learning from breaking pseudorandom function families}\label{s:prfs}

Oliveira and Santhanam \cite{OS} showed that the assumption of the existence of natural proofs from Theorem \ref{t:cikk} can be further weakened to the existence of a distinguisher breaking non-uniform pseudorandom function families. Their result follows from a combination of Theorem \ref{t:cikk} and the Min-Max Theorem. Using their strategy but combining the Min-Max Theorem with Theorem \ref{t:gcsplear}, learning algorithms with random examples can be obtained from distinguishers breaking succinct non-uniform pseudorandom function families
\bigskip

A {\em two-player zero-sum game} is specified by an $r\times c$ matrix $M$ and is played as follows. MIN, the row player, chooses a probability distribution $p$ over the rows. MAX, the column player, chooses a probability distribution $q$ over the columns. A row $i$ and a column $j$ are drawn randomly from $p$ and $q$, and MIN pays $M_{i,j}$ to MAX. MIN plays to minimize the expected payment, MAX plays to maximize it. The rows and columns are called the {\em pure strategies} available to MIN and MAX, respectively, while the possible choices of $p$ and $q$ are called {\em mixed strategies}. The Min-Max theorem states that playing first and revealing one's mixed strategy is not a disadvantage: 
$$min_{p} max_j \sum_i p(i)M_{i,j}=max_q min_i \sum_j q(j) M_{i,j}.$$ Note that the second player need not play a mixed strategy - once the first player's strategy is fixed, the expected payoff is optimized for the second player by playing some pure strategy. The expected payoff when both players play optimally is called the {\em value} of the game. We denote it $v(M)$.

A mixed strategy is {\em $k$-uniform} if it chooses uniformly from a multiset of $k$ pure strategies. Let $M_{min}=min_{i,j} M_{i,j}$ and $M_{max}=max_{i,j} M_{i,j}$. Newman \cite{Nm}, Alth\"ofer \cite{Alt} and Lipton-Young \cite{LY} showed that each player has a near-optimal $k$-uniform strategy for $k$ proportional to the logarithm of the number of pure strategies available to the opponent.

\begin{theorem}[\cite{Nm, Alt, LY}]\label{thm:minmax} For each $\epsilon>0$ and $k\ge \ln(c)/2\epsilon^2$, $$min_{p\in P_k} max_j \sum_{i} p(i) M_{i,j}\le v(M)+\epsilon(M_{max}-M_{min}),$$ where $P_k$ denotes the $k$-uniform strategies for MIN. The symmetric result holds for MAX.
\end{theorem}

\begin{definition}[Succinct non-uniform PRF] An $(m,m')$-succinct non-uniform pseudorandom function family from circuit class $\mathcal{C}$ safe against circuits of size $s$ is a set $S$ of partial truth-tables $\langle (x_1,b_1),\dots, (x_m,b_m)\rangle$ where each $x_i$ is an $n$-bit string and $b_i\in\{0,1\}$ such that each partial truth-table from $S$ is computable by one of $m'$ circuits from $\mathcal{C}$ and for every circuit $D$ of size $s$, $$\Pr_{x}[D(x)=1]-\Pr_{x\in S}[D(x)=1]<1/s$$ where the first probability is taken over $x\in\{0,1\}^{m(n+1)}$ chosen uniformly at random and the second probability over partial truth-tables chosen uniformly at random from $S$.
\end{definition}

\begin{theorem}[Learning or succinct non-uniform PRF]\label{t:towardscore}
Let $c\ge 1$ and $s>n,m\ge 1$. There is an $(m,8s^4)$-succinct non-uniform PRF in $\Circuit[n^c]$ safe against $\Circuit[s]$ or there are circuits of size $poly(s)$ learning $\Circuit[n^c]$ over the uniform distribution with random examples, confidence $1/poly(s)$, up to error $1/2-1/poly(s)$.
\end{theorem}

\proof 

Consider a two-player zero-sum game specified by a matrix $M$ with rows indexed by $n^c$-size circuits with $n$ inputs and columns indexed by $s$-size circuits with $m(n+1)$ inputs. Define the entry $M_{C,D}$ of $M$ corresponding to a row circuit $C$ and a column circuit $D$ as $$M_{C,D}:=|\Pr_x[D(x)=1]-\Pr_x[D(G_C(x))=1]|$$ for the generator $G_C$ from the proof of Lemma \ref{l:gen}. Hence $M_{max}-M_{min}\le 1$. 

If $v(M)\ge 1/4s$, then by Theorem \ref{thm:minmax} (with $\epsilon=1/8s$), there exist a multiset of $k\le 32n^{c+1}s^2$ $s$-size circuits $D^1,\dots, D^{k}$ such that for every $n^c$-size circuit $C$, a random $D$ from $D^1,\dots, D^{k}$ satisfies $$\text{E}[|\Pr[D(x)=1]-\Pr[D(G_C(x))=1]|]\ge 1/8s.$$ 

By Lemma \ref{l:gen}, for every $n^c$-size circuit $C$, one of the circuits $D^1,\dots, D^k$ (or their negations) can be used to learn $C$ with confidence $1/poly(s)$, up to error $1/2-1/poly(s)$. 
A $poly(s)$-size circuit using a random $D^i$ from $D^1,\dots, D^k$ or its negation thus learns $\Circuit[n^c]$ with random examples, confidence $1/poly(s)$, up to error $1/2-1/poly(s)$.
\medskip

If $v(M)<1/4s$, then by Theorem \ref{thm:minmax} (with $\epsilon=1/4s$), there exists a multiset of $k\le 8s^4$ $n^c$-size circuits $C^1,\dots, C^k$ such that for every $s$-size circuit $D$, a random $C$ from $C^1,\dots, C^k$ satisfies $$\text{E}[|\Pr[D(x)=1]-\Pr[D(G_C(x))=1]|]\le 1/2s.$$ Since $\text{E}[|\Pr[D(x)=1]-\Pr[D(G_C(x))=1]|]\ge|\Pr[D(x)=1]-\text{E}[\Pr[D(G_C(x))=1]]|$ a generator $$G:\{0,1\}^{mn+\lceil\log k\rceil}\mapsto \{0,1\}^{mn+m}$$ which takes as input a string of length $mn+\lceil\log k\rceil$ encoding (an index of) a circuit $C$ from $C^1,\dots, C^k$ together with $m$ $n$-bit strings $x_1,\dots,x_m$ and outputs $x_1,C(x_1),\dots, x_m, C(x_m)$ is safe against circuits of size $s$. The range of $G$ defines an $(m,8s^4)$-succinct non-uniform PRF in $\Circuit[n^c]$ safe against $\Circuit[s]$.
\qed

\bigskip
Note that the existence of a generator $G$ from the proof of Theorem \ref{t:towardscore} follows directy from a counting argument if we do not require that $G$ defines a PRF of small complexity: a random set of $poly(s,n)$ strings (yielding a non-uniform pseudorandom generator mapping $\{0,1\}^{O(\log s)}$ to $\{0,1\}^n$) fools circuits of size $s$. 

\def\unimin{
\bigskip
\noindent {\bf Lowering the bar to complexity-theoretic PRGs.} Let $M^{n,m}$ be a sequence of $2^n\times 2^s$ matrices specifying two-player zero-sum games. We say that a $poly(s)$-time algorithm $A$ is a sampling algorithm of a mixed strategy $q$ of the column player if $\forall j\in [2^s]$, $\Pr_{x\in\{0,1\}^{n}}[A(x)=j]=q(j)$.

\begin{hypothesis}[Uniform Min-Max]\label{h:uniminmax} Let $M^{n,s}$ be a sequence of $2^n\times 2^s$ matrices specifying two-player zero-sum games. Assume there is a p-time algorithm which given a sampling algorithm of a mixed strategy $q$ of the column player outputs a pure strategy $i$ of the row player such that $\sum_j q(j)M_{i,j}\le v(M)$. Then there is a p-time algorithm which given $1^s$ outputs a sampling algorithm of a mixed strategy $q'$ of the column player such that $min_i \sum_j q'(j)M_{i,j}\ge v(M)-1/2^s$.
\end{hypothesis}

A uniform version of the Min-Max theorem has been proved by Vadhan-Zheng \cite{VZ}. However, their algorithm does not seem to generate sampling algorithms of mixed strategies - just an implicit description of such strategies.

\begin{theorem}[Non-refutation of Learning or PRGs assuming uniform Min-Max]\label{t:almostcore}
Assume the uniform Min-Max hypothesis. Then, there is a p-time computable pseudorandom generator safe against $\Circuit[s]$, where $s\ge poly(n)$, or for every constant $c$, there is no p-time algorithm which given a circuit $D$ of size $poly(s,n)$ outputs a circuit $C$ of size $n^c$ such that $D$ does not learn $C$ with random examples, confidence $1/n$, up to error $1/2-1/poly(s,n)$. 
\end{theorem}

\proof Proceed as in the proof of Theorem \ref{t:towardscore} but distinguish the following two cases. If for infinitely many $n$, $v(M)\ge 1/4s(n)$, then for infinitely many $n$, there are circuits of size $poly(s,n)$ learning circuits of size $n^c$ with random examples, confidence $1/n$, up to error $1/2-1/poly(s,n)$. If for all sufficiently big $n$, $v(M)<1/4s(n)$, then apply uniform Min-Max hypothesis to conclude that for all sufficiently big $n$, for every circuit $D$ of size $s(n)$, the generator $$G:\{0,1\}^{mn+poly(s)}\mapsto \{0,1\}^{mn+m}$$ which takes as input a string of length $mn+poly(m)$ encoding an input of the sampling algorithm from the uniform Min-Max hypothesis together with $m$ $n$-bit strings $x_1,\dots,x_m$ and outputs $x_1,C(x_1),\dots, x_m, C(x_m)$ is safe against circuits of size $s$. \qed

\bigskip

Since the existence of an efficient algorithm generating strategies for the second Player follows from the provability of a suitable statement in intuitionistic \SB we can instead assume that the non-existence of an efficient learner is also feasibly provable.

\begin{corollary}[Consistency of Learning or PRGs assuming uniform Min-Max]\label{c:conscore}
Assume uniform Min-Max hypothesis. Then, there is a p-time computable pseudorandom generator safe against $\Circuit[s]$, where $s\ge poly(n)$, or for every constant $c$, it is consistent with intuitionistic $\SB$ that there are circuits of size $poly(s,n)$ learning $\Circuit[n^c]$ with random examples, confidence $1/n$, up to error $1/2-1/poly(s,n)$. 
\end{corollary}

\proof[Proof sketch.] Proceed as in the proof of Theorem \ref{t:towardscore} but in the case $v(M)<1/4s$ apply witnessing theorem. \qed
}

\subsection{Superbits vs demibits}\label{s:rudich}

Rudich \cite{Rs} proposed a conjecture about the existence of superbits, a version of pseudorandom generators safe against nondeterministic circuits, and showed that it rules out the existence of \NP-natural properties against \Ppoly. He then asked whether the existence of superbits follows from a seemingly weaker assumption of the existence of so called demibits. We note that an affirmative answer to his question would resolve Question \ref{q:dichotomy} in nondeterministic setting.

\begin{definition}[Superbit] A function $g:\{0,1\}^n\mapsto\{0,1\}^{n+1}$ computable by p-size circuits is a superbit if there is $\epsilon<1$ such that for infinitely many input lengths $n$, for all nondeterministic circuits $C$ of size $|C|\le 2^{n^{\epsilon}}$, $$\Pr_{x\in\{0,1\}^{n+1}}[C(x)=1]-\Pr_{x\in\{0,1\}^n}[C(g(x))=1]<1/|C|.$$
\end{definition}

\begin{definition}[Demibit] A function $g:\{0,1\}^n\mapsto\{0,1\}^{n+1}$ computable by p-size circuits is a demibit if there is $\epsilon<1$ such that for infinitely many input lengths $n$, no nondeterministic circuit $C$ of size $|C|\le 2^{n^{\epsilon}}$ satisfies $$\Pr_{x\in\{0,1\}^{n+1}}[C(x)=1]\ge 1/|C|\ \ \ \text{ and }\ \ \ \Pr_{x\in\{0,1\}^n}[C(g(x))=1]=0.$$
\end{definition}

\begin{proposition}[Question \ref{q:dichotomy} vs Rudich's problem] Assume the existence of demibits implies the existence of superbits. Then, either superbits exist or for each $c\ge 1$, for each $\epsilon<1$, $\Circuit[n^c]$ is learnable by $\Circuit[2^{O(n^{\epsilon})}]$ over the uniform distribution with random examples, confidence $1/2^{O(n^{\epsilon})}$ up to error $1/2-1/2^{O(n^{\epsilon})}$, where the learner is allowed to generate a nondeterministic or co-nondeterministic circuit approximating the target function.
\end{proposition}

\proof Assume superbits do not exist and their non-existence implies the non-existence of demibits. Consider a generator $G:\{0,1\}^{mn+n^{c+1}}\mapsto\{0,1\}^{mn+m}$, with $m=n^{c+1}+1$, which interprets the first $n^{c+1}$ bits of its input as a description of an $n^c$-size circuit $C$ and then computes on the remaining $mn$ inputs as generator $G_C$ from Lemma \ref{l:gen}. Since $G$ is not a demibit, for each $\epsilon<1$ there are nondeterministic circuits $D$ of size $2^{(mn+m-1)^{\epsilon}}$, such that for each $n^c$-size circuit $C$, $$\Pr[D(x)=1]-\Pr[D(G_C(x))=1]\ge 1/|D|.$$ By the proof of Lemma \ref{l:gen}, this means that $n^c$-size circuits are learnable by circuits of size $poly(|D|)$ with confidence $1/poly(|D|)$ up to error $1/2-1/poly(|D|)$, except that the learner might generate nondeterministic (if $r_i=0$) or co-nondeterminitic (if $r_i=1$) circuit approximating the target function. \qed

\section{Learning speedup}\label{s:speedup}

A striking consequence of the relation between natural proofs and learning algorithms is a learning speedup of Oliveira and Santhanam \cite{OS}. 

Suppose \Ppoly is learnable by circuits of weakly subexpoential size $2^n/n^{\omega(1)}$. The learning circuits can be used to accept truth-tables of all functions in \Ppoly while their size guarantees that many hard functions are going to be rejected. This implies the existence of a \Ppoly-natural property useful against \Ppoly, which by Theorem \ref{t:cikk}, gives us circuits of strongly subexponential size $2^{n^\gamma}$, $\gamma<1$, learning \Ppoly.

The argument of Oliveira and Santhanam can be generalized to a speedup of learners of arbitrary size $s$. 
Here, we show how to derive such a generalized version more directly without constructing natural proofs and invoking Theorem \ref{t:cikk}. This is possible thanks to a more direct exploitation of a slightly modified NW-generator. A drawback of the approach is that we need to assume learning with random examples instead of membership queries.

\begin{theorem}[Generalized speedup]\label{t:speedup} Let $d,k\ge 1$ and $n\le s(n)\le 2^{n}/n$. Assume $\Circuit[n^{10dk}]$ is learnable by $\Circuit[s(n)]$ over the uniform distribution with random examples, confidence $1$, up to error $1/2-5/n$. Then circuits of size $m^k$ with $m=n^d$ inputs are learnable by circuits of size $n^{dK}(s(n))^3$ over the uniform distribution with non-adaptive membership queries, confidence $1/n^3$, up to error $1/2-1/n$. Here, $K$ is an absolute constant.
\end{theorem}


Theorem \ref{t:speedup} implies, for example, that if p-size circuits are learnable with random examples by circuits of quasipolynomial size $n^{O(\log n)}$, then p-size circuits are learnable with membership queries by circuits of size $O(n^{\epsilon \log n})$, for each $\epsilon>0$. The speedup is achieved w.r.t. the input length of target functions at the expense of their circuit complexity.

\proof 
Let $A$ be a $2^{b}\times u$ 0-1 matrix forming a $(b,n^{d})$-design with $|J_i(A)|=n^{d}$ for $n^{2d}\le u\le 2n^{2d}$, a constant $d$ and parameter $b$ such that $ns\le 2^{b}\le 2ns$. The design is constructed in the usual way by evaluating polynomials of degree $\le b$ on $n^{d}$ points of a field with $n^d\le p\le 2n^d$ elements. 
In particular, there are $n^{9d}$-size circuits which given $i\in\{0,1\}^{b}$ and $w\in\{0,1\}^u$ output $w|J_i(A)$. Define $NW_f$-generator mapping strings $w$ of length $u$ to strings of length $2^{n}$ as $$(NW_f(w))_{x_1,\dots,x_n}=f(w|J_{x_1,\dots,x_{b}}(A)).$$ Then for each $m$-input function $f\in\Circuit[m^k]$ and $w\in\{0,1\}^{u}$, $(NW_f(w))_x$ is computable as a function of $x\in\{0,1\}^n$ by a circuit of size $n^{10dk}$. 

By the assumption of the theorem every such circuit $(NW_f(w))_x$ is learnable by a circuit $L$ of size $s$ with confidence $\delta=1$, up to error $1/2-\epsilon$. Consequently, there is a circuit $D^f$ of size $O(s^3)$ such that \begin{equation}\label{e:speed1}\Pr_{w,x,y^1,\dots,y^t}[D^f(x_1,\dots,x_n,w,y^1,\dots,y^t)=f(w|J_{x_1,\dots,x_{b}}(A))]\ge (1/2+\epsilon)\delta\end{equation} where $D^f$ queries values $f(w|J_{y^j}(A))$ for $t\le s$ random strings $y^j\in\{0,1\}^{b}$, $j=1,\dots, t$. The size of $D^f$ takes into account the need to simulate the circuit described by $L$. Now, random $y^1,\dots,y^t$ satisfy \begin{equation}\label{e:speed3}\Pr_{w,x}[D^f(x_1,\dots,x_n,w,y^1,\dots,y^t)=f(w|J_{x_1,\dots,x_{b}}(A))]\ge 1/2+\epsilon-1/n\end{equation} with probability at least $1/n$. Otherwise, the probability in (\ref{e:speed1}) would be $<1/n+(1/2+\epsilon-1/n)$. Similarly, given $y^1,\dots,y^t$ such that (\ref{e:speed3}) holds, a random $x\in\{0,1\}^n$ 
satisfies \begin{equation}\label{e:speed2}\Pr_{w}[D^f(x_1,\dots,x_n,w,y^1,\dots,y^t)=f(w|J_{x_1,\dots,x_{b}}(A))]\ge 1/2+\epsilon-3/n\end{equation} with probability at least $2/n$. 
Moreover, since every $y^j$ specifies $2^{n-b}$ values of $(NW_f(w))_x$, given $y^1,\dots, y^t$, a random $x\in\{0,1\}^n$ equals some $y^j$ on the first $b$ bits with probability $\le t/2^b\le 1/n$. Applying the same averaging one more time, for $y^1,\dots,y^t$ and $x$ which differs on the first $b$ bits from each $y^j$ and satisfies (\ref{e:speed2}), randomly fixed $u-n^d$ bits of $w$ on the positions of $[u]\backslash J_{x}(A)$ preserve the probability (\ref{e:speed2}) up to an additional error $1/n$ with probability at least $1/n$.

For each $y^1,\dots,y^t$, each $x$ which differs on the first $b$ bits from every $y^j$ and for each fixation of $u-n^d$ bits of $w$ on the positions of $[u]\backslash J_{x}(A)$, $(b,n^d)$-design guarantees that the number of all queries $f(w|J_{y^j}(A))$, $j=1,\dots,t$, of $D^f$ for all possible $w$ with the $u-n^d$ fixed bits is $\le t2^{b}$. We can thus learn  a circuit $D'$ approximating $f\in \Circuit[m^k]$ with $m=n^d$ inputs with advantage $1/2+\epsilon-4/n$ in the following way. Choose random $y^1,\dots,y^t$, $x$, random $u-n^d$ bits of $w$ corresponding to $[u]\backslash J_{x}(A)$ and query $\le t2^b$ values $f(w|J_{y^j}(A))$ for all possible $w$ with the $u-n^d$ fixed bits. Then the circuit $D'$, given $n^d$ bits of $w$ corresponding to $J_x(A)$, generates $w$ and computes as $D^f$ with the provided queries $f(w|J_{y^j}(A))$. Since $w$ can be constructed from given $n^d$ bits, $x$ and the $u-n^d$ fixed bits of $w$ by a circuit of size $n^{O(d)}$, each $w|J_{y^j}(A)$ can be constructed from $w$ and $y^j$ by a circuit of size $n^{9d}$ and for each query to $f$ the right value can be selected by a circuit of size $O(n^dt2^b)$, the size of $D'$ is $O(s^3+tn^{9d}+n^dt^22^b+n^{O(d)})\le n^{O(d)}s^3$. $D'$ can be described by $n^{dK}s^3$ bits, for an absolute constant $K$, and constructed by a circuit of the same size which just substitutes $y^j, x$ and $u-n^d$ bits of $w$ in the otherwise fixed description of $D'$.

Since random $y^1,\dots,y^t$ satisfy (\ref{e:speed3}) with probability at least $1/n$, a random $x$ differs on the first $b$ bits from each $y^1,\dots,y^t$ and satisfies (\ref{e:speed2}) with probability at least $1/n$ while the randomly fixed $u-n^d$ bits of $w$ have the desired property with probability at least $1/n$ as well, the confidence of the learning algorithm is at least $1/n^3$. \qed

\bigskip

We give one more proof of the learning speedup which also addresses the issue of membership queries.

\begin{theorem}[Alternative speedup]\label{t:aspeed}
Let $d\ge 2; k\ge 1$ and 
$\epsilon<1$. Assume $\Circuit[n^{10dk}]$ is learnable by $\Circuit[2^{\epsilon n}]$ over the uniform distribution (possibly with membership queries) with confidence $1$, up to error $1/n^5$. Then, circuits of size $n^{dk}$ with $n^d$ inputs are learnable by circuits of size $2^{Kn}$ over the uniform distribution with confidence $1/2^{Kn}$ up to error $1/2-2^{Kn}$, where $K$ is an absolute constant.
\end{theorem}

\proof By a counting argument there exists $H$ which is not $(1-1/n)$-approximable by circuits of size $2^{\epsilon n}$. Here, $n$ is w.l.o.g. sufficiently big. By Lemma \ref{l:mainconverse}, learnability of $\Circuit[n^{10dk}]$ by $\Circuit[2^{\epsilon n}]$ up to error $1/n^5$ implies the existence of circuits of size $2^{O(n)}$ witnessing errors of circuits of size $n^{10dk}$ with probability $\ge 1-2/n^4$. The conclusion thus follows by applying Theorem \ref{t:main}. The improved confidence and approximation parameter is the consequence of the fact that our witnessing circuits succeed in the first round, i.e. $t=1$. \qed
\bigskip

\noindent {\bf Proof-search speedup.} The core trick behind Theorem \ref{t:speedup} can be formulated in the context of 
proof complexity. Assume that an $n^{10dk}$-size lower bound is provable in a proof system $P$ by a proof of size $s(n)$. Then, a substitutional instance of the same $P$-proof of size $s(n)$ proves an $m^k$-size lower bound for circuits with $m=n^d$ inputs, on inputs given by the NW-generator from the proof of Theorem \ref{t:speedup}. Here, the base function of the NW-generator is not specified but represented by free variables encoding a circuit of size $m^k$.

\def\prfspeed{The observation can be formulated also more generally as a form of a proof-search speedup. In fact, in order to do so, we do not need the modified NW-generator from Theorem \ref{t:speedup}, the standard one will suffice.
\smallskip

Denote by $\lb_Y(f,t)$ a $poly(t,n)$-size propositional formula expressing a $t(n)$-size circuit lower bound for a Boolean function $f$ restricted to inputs 
$Y\subseteq\{0,1\}^n$. 
The formula $\lb_Y(f,t)$ has $poly(t)$ variables representing circuits of size $t$ and contains also auxiliary variables needed to encode computations of $t$-size circuits on inputs from $Y$. The auxiliary variables are not essential in the sense that their true value can be determined in p-time from $Y$ and an assignment encoding a $t$-size circuit.  
If $Y=\{0,1\}^n$, we denote the resulting formula $\lb_Y(f,t)$ by $\ttable(f,t)$. 
Further, we say that a proof system $P$ admits substitutions if for each $s$-size $P$-proof of a formula $\phi$ and a partial assignment $a$ of variables of $\phi$, we can find a $P$-proof of $\phi(a)$ in time $O(s)$.

\begin{theorem}[Proof-search speedup]\label{t:autsp} Let $d,k\ge 1$. Assume that a proof system $P$ admits subtitutions and there are $2^{O(n)}$-size circuits which given a tautology $\ttable(f,n^{5dk})$ find its $P$-proof. Then there are $2^{O(n)}$-size circuits which find a $P$-proof of each tautology $\lb_Y(h,m^k)$, where $h$ is a Boolean function with $m=n^d$ inputs which is hard for circuits of size $m^{Kdk}$, for a sufficiently big absolute constant $K$, $Y$ is a set of $n^d$-bit strings of the form $w|J_i(A)$ for a $2^n\times u$ matrix $A$ being the $(n,n^d)$-design from the proof of Theorem \ref{t:main} and $w\in\{0,1\}^u$.
\end{theorem}

Theorem \ref{t:autsp} shows that if we can find proofs of p-size lower bounds for Boolean functions with $n$ inputs by circuits of size $2^{O(n)}$, then we can find proofs of p-size lower bounds for Boolean functions with $n$ inputs satisfying the conditions of Theorem \ref{t:autsp} by circuits of size $2^{O(n^\epsilon)}$, for each $\epsilon>0$. 

\proof Given a tautology $\lb_Y(h,m^k)$ we consider a formula $\ttable(f,n^{5dk})$ such that $\lb_Y(h,m^k)$ can be obtained from $\ttable(f,n^{5dk})$ by a partial assignment of its variables. Such a formula with a function $f$ exists and is specified by the assumption of the theorem.

The formula $\ttable(f,n^{Kdk})$ is a tautology as well. Otherwise, there would be a circuit of size $n^{5dk}$ computing $f$ which could be used to compute $h$: The set $Y$ consists of strings $w|J_i(A)$. By the construction of design $A$, each $w|J_i(A)$ specifies $n^d$ points of a polynomial of degree $n$. Given $w$ for each $w|J_i(A)$ we can determine a polynomial $p$ of degree $n$ which generates $w|J_i(A)$. In fact, there are $poly(n)$-size circuits which given $w$ and $w|J_i(A)$ output the polynomial, represented as a sum of monomials, together with the index $i$ (PROBLEM: not index i but some polynomial). Composing these circuits with $C$ yields a circuit of size $m^{Kdk}$ computing $h$. This specifies the constant $K$ and contradicting the hardness of $h$. 

Therefore, by the assumption of the theorem we can find a $P$-proof of $\ttable(f,n^{5dk})$ using circuits of size $2^{O(n)}$. As $P$ admits substitutions and there are circuits constructing the design $A$ from the proof of Theorem \ref{t:main}, we obtain also a $2^{O(n)}$-size $P$-proof of $\lb_Y(h,m^k)$. \qed
}

\bigskip

\noindent {\bf Nonlocalizable hardness magnification.} Theorem \ref{t:speedup} and the original speedup of Oliveira and Santhanam can be interpreted as hardness magnification theorems. Hardness magnification is an approach to strong complexity lower bounds by reducing them to seemingly much weaker lower bounds developed in a series of recent papers \cite{HM, MP, OPS, MMW, CMMW, CT, CJWs, CHOPRS, CJWt, Msca, CHMY}, see \cite{CHOPRS} for a more comprehensive survey. For example, it turns out that in order to prove that functions computable in nondeterministic quasipolynomial-time are hard for \NC it suffices to show that a parameterized version of the minimum circuit size problem \MCSP is hard for $\AC[2]$. However, \cite{CHOPRS} identified a {\em locality barrier} which explains why direct adaptations of many existing lower bounds do not yield strong complexity lower bounds via hardness magnification. Essentially, the reason is that the existing lower bounds for explicit Boolean functions work often even for models which are allowed to use arbitrary oracles with $n^{o(1)}$-small fan-in. This is easy to see in the case of $\AC[2]$ lower bounds: oracles of small fan-in can be simulated by polynomials of low degree. On the other hand, hardness magnification theorems typically yield (unconditional) upper bounds in the form of weak computational models extended with local oracles computing specific problems such as the abovementioned version of \MCSP. In fact, even irrespective of hardness magnification it is important to develop lower bound methods which do not localize: proving the nonexistence of subexponential-size learning algorithms for \Ppoly would imply the nonexistence of \Ppoly natural properties against \Ppoly but it is not hard to see that natural properties against \Ppoly are computable by p-size circuits with local oracles. Overcoming the locality barrier is thus essential for proving strong complexity lower bounds in general.\footnote{Some known circuit lower bounds above the magnification threshold are provably nonlocalizable but they do not fit to the framework of the so called Hardness Magnification frontier \cite{CHOPRS}, one reason being that they do not work for explicit and natural problems, cf. \cite{CHOPRS, CJWt}. For example, a nonlocalizable lower bound from \cite{CHOPRS} works for a function in $\mathsf{E}$ which is artificial in the sense that it is designed to avoid localization, not for a problem of independent interest  such as \MCSP.  Oliveira \cite{Okol} showed that near superlinear-size lower bounds for a version of \MCSP defined w.r.t. a notion of randomized Kolmogorov complexity imply strong circuit lower bounds while the same problem is provably hard for probabilistic p-time. The lower bound of Oliveira works, however, only against uniform models of computation. Moreover, the magnification theorem concludes at best a `weak' lower bound of the form quasipolynomial-time $\mathsf{QP}$ being hard for \Ppoly. Similarly, an approach of Chen, Jin and Williams \cite{CJWt} via derandomizations and uniform obstructions appears to avoid the locality barrier but yields at best lower bounds of the form $\mathsf{QP}\not\subseteq\Ppoly$.}

Theorem \ref{t:speedup}, if read counterpositively, is a magnification of $O(n^{\epsilon \log n})$-size lower bounds for learning p-size circuits to $n^{O(\log n)}$-size lower bounds. This differs from previous hardness magnification theorems 
by avoiding localization: the size of the learner plays a crucial role in the reduction and therefore cannot be simply replaced by an arbitrary oracle. The same trick is behind non-blackbox worst-case to average-case reductions within \NP of Hirahara \cite{Hb}. To the best of my knowledge, the only other hardness magnification theorems with this property 
appeared in \cite{CHOPRS} and \cite{Hmeta}.\footnote{There are two more results which could be potentially classified as nonlocalizable hardness magnifications. A theorem of Buresh-Oppenheim and Santhanam \cite[Theorem 1]{JOS} is based on an exploitation of Nisan-Wigderson generators similar to that of \cite{CHOPRS} but it seems less practical in its current form, as it magnifies only lower bounds for nondeterministic circuits. The other result of Tal \cite{Tmag} shows that an average-case hardness for formulas of size $s$ can be magnified to the worst-case hardness for slightly bigger formulas. A problem is that \cite{Tmag} magnifies at best to an $s^2$-size lower bound. Moreover, if we wanted to strenghten it further by connecting it with another magnification theorem, it is not clear how to preserve the nonlocalizability - the weak lower bound obtained via \cite{Tmag} would likely localize.} \cite[Theorem 1]{CHOPRS}, like Hirahara \cite{Hb} and the speedup of Oliveira-Santhanam, is based on the result of Carmosino, Impagliazzo, Kabanets and Kolokolova \cite{CIKK}. However, the hardness magnification from \cite{CHOPRS} is still captured by the locality barrier: it asks for a lower bound for a version of \MCSP whose localized version does not hold (as witnessed by other hardness magnification theorems). Theorem \ref{t:speedup} does not seem to localize in this sense either: it asks for an $n^{\epsilon \log n}$-size lower bound on learning algorithms while there seems to be no reason to expect that p-size circuits are learnable by circuits of size $O(n^{\log n})$ extended with oracles of fan-in $n^{o(1)}$. (Such a localization would mean that p-size circuits are learnable in subexponential size.) 
The magnification theorems of Hirahara \cite{Hmeta} face similar complications.\footnote{Hirahara \cite[Theorem 11 and 13]{Hmeta} proves two types of magnification theorems. The first type essentially adapts the result from \cite{CHOPRS} in the context of weaker computational models. The second type extends it by introducing metacomputational circuit lower bound problems MCLPs and showing that weak lower bounds for MCLPs can be magnified as well. MCLPs are not solvable by any algorithm whatsoever unless standard hardness assumptions break. This implies that there is no unconditional upper bound for MCLPs and the locality barrier does not apply. Unfortunately, we do not have any interesting lower bound for MCLPs either. The corresponding magnification theorems thus do not establish a Hardness Magnification frontier~\cite{CHOPRS}. Nevertheless, as suggested in \cite{Hmeta}, developing such methods might be a way to strong lower bounds.}

Unfortunately, Theorem \ref{t:speedup} does not reduce p-size lower bounds to, say, subquadratic lower bounds: It magnifies $n^{O(d)}s^3$-size lower bounds for learning functions with $m=n^d$ inputs (and circuit complexity $m^{k}$) to an $s$-size lower bound for learning functions with $n$ inputs (and circuit complexity $n^{10dk}$). That is, a polynomial speedup w.r.t. the input-length of target functions is traded for a polynomial decrease of the circuit size of target functions. Ideally, we would like to magnify, say, $n^{1.9}$-size formula lower bound for learning circuits of size $n^{1.1}$ with $n$ inputs to $n^{O(1)}$-size formula lower bounds for learning circuits of size $n^{2.1}$ with $n$ inputs. If the existing methods for proving the required formula lower bounds were applicable to prove subquadratic formula lower bounds for learning algorithms (note that such lower bounds are allowed to localize and naturalize), such a strengthening of Theorem \ref{t:speedup} would lead to explicit \NC lower bounds.

\section{Concluding remarks and open problems}\label{s:concluding}

The methods for deriving learning algorithms from circuit lower bounds presented in this paper might be improvable in many ways.
\medskip

\noindent {\bf Safe cryptography or efficient learning.} Perhaps the most appealing question asks for bridging cryptography and learning theory. Showing that efficient learning follows from breaking pseudorandom generators, i.e. answering positively Question \ref{q:dichotomy}, would establish a remarkable win-win situation. As discussed in Section \ref{s:rudich} the question is closely related to a problem of Rudich about turning demibits to superbits. 
\medskip

\noindent{\bf Instance-specific learning vs PAC learning.} Circuit lower bounds correspond to a simple instance-specifc learning model described in Section \ref{s:i-s}. Can we improve our understand of the model and its relation to PAC learning? In particular, can we determine how much we can learn from a single circuit lower bound? A possible formalization of the problem is given by Question \ref{q:fip}. 
\medskip

\noindent {\bf Connections to proof complexity.} The present paper brings several methods from proof complexity to learning theory. It seems likely that these connections can be strengthened. A particularly relevant part of proof complexity is the theory of proof complexity generators, cf. \cite{Kfor}. An interesting conjecture in the area due to Razborov \cite{Rkdnf} implies a conditional hardness of circuit lower bounds in strong proof systems. In other words, Razborov's conjecture asks for turning short proofs of circuit lower bounds into upper bounds breaking standard hardness assumptions. 

Notably, strengthening Theorem \ref{t:main} by allowing white-box access in the witnessing of lower bounds would lead to a conditional unprovability of p-size lower bounds for $\SAT$ in Cook's theory \PV. A complication is that under standard hardness assumptions such a witnessing exists. That is, in order to obtain the conditional unprovability, one might need to exploit the \PV-provability in a deeper way. Nevertheless, this suggests a simplified version of Question \ref{q:dichotomy}: Can we prove a disjunction stating the \PV-consistency of the existence of strong pseudorandom generators or the \PV-consistency of efficient learning? Since, by witnessing theorems in \PV, both the \PV-provability of the non-existince of pseudorandom generators and the \PV-provability of the impossibility of effficient learning imply uniform efficient algorithms witnessing these facts, it could be possible to combine them with a version of uniform MinMax \cite{VZ} to get a contradiction. 
\medskip


\noindent {\bf Nonlocalizable hardness magnification near the existing lower bounds.} Can we push forward the program of hardnness magnification by strengthening the magnification from Theorem \ref{t:speedup} to a setting in which strong circuit lower bounds follow from lower bounds near the already existing ones? The importance of the question stems from the necessity of developing nonlocalizable magnification theorems or nonlocalizable constructive lower bound methods as discussed in Section \ref{s:speedup}.
\medskip

\noindent {\bf SAT solving circuit lower bounds.} It would be interesting to investigate practical consequences of the provability of circuit lower bounds. Circuit lower bounds for explicitly given Boolean functions are \coNP statements which means that they are encodable into propositional tautologies resp. SAT instances. Could SAT solvers be successful in proving interesting instances of circuit lower bounds for some fixed input lengths? If so, this could provide an experimental verification of central results and conjectures from complexity theory such as $\Ptime\ne\NP$ up to some finite domain. As discussed in the present paper, efficient algorithms proving circuit lower bounds can be also transformed into learning algorithms, which provides a separate motivation for this line of research.

In particular, SAT solving of circuit lower bounds could lead to an interesting comparison with the research on neural networks. The task of training a neural network is to design a circuit $C$ of size $s$, typically with a specific architecture, coinciding with some training input samples $(y_i,f(y_i))$, and apply it to predict the value $f(y)$ on a new input $y$. As discussed in Section \ref{s:i-s}, this problem can be addressed by proving a circuit lower bound. Since proving a circuit lower bound can give us a reliable instance-specific prediction one could try to use SAT solvers to verify outcomes of neural networks. More generally, one could try to simulate neural networks by SAT solving circuit lower bounds. A potential advantage of SAT solvers is that they do not need to construct a circuit coinciding with training data - it is enough to prove its properties (lower bounds). On the othe hand, SAT solvers need to prove a universal statement which might turn out to be even harder. 

\def\practicalpart{
\section{SAT solving circuit lower bounds}

Circuit lower bounds for explicitly given Boolean functions are \coNP statements which makes them easily encodable into propositional tautologies resp. SAT instances. Could SAT solvers be successful in proving interesting instances of circuit lower bounds for some fixed input lengths? If so, this could provide an experimental verification of central results and conjectures from complexity theory such as $\Ptime\ne\NP$ up to some finite domain. Efficient algorithms proving circuit lower bounds can be also transformed into learning algorithms, which provides a separate motivation for this line of research.

\subsection{SAT solvers versus neural networks}

\noindent The task of training a neural network is to design a circuit $C$ of size $s$, typically with a specific architecture, coinciding with some training input samples $(y_i,f(y_i))$, and apply it to predict the value $f(y)$ on a new input $y$. As discussed in Section \ref{seclear}, this problem can be addressed by proving a circuit lower bound of the form:  $$\forall\text{ circuit }C\text{ of size }s,\ \bigvee_{i} C(y_i)\ne f(y_i)\vee C(y)\ne\epsilon$$ for $\epsilon\in\{0,1\}$. If the lower bound holds, then either there is no neural network of size $s$ for given inputs $(y_i,f(y_i))$ or the right prediction on $y$ is $1-\epsilon$. 
\smallskip

\noindent {\bf $\circ$ verifying neural networks with SAT solvers.} Proving a circuit lower bound can thus guarantee that no neural network predicts $\epsilon$ as the outcome. 
\smallskip

\noindent {\bf $\circ$ simulating neural networks with SAT solvers.} Could SAT solving circuit lower bounds be easier than learning neural networks? A potential advantage of SAT solvers is that they do not need to construct a circuit coinciding with training data - it is enough to prove its properties (lower bounds). On the othe hand, SAT solvers need to prove a universal statement which might turn out to be even harder.

\subsection{Problem statement}

The problem is to decide the existence of circuits with size $s$ for $n$-input 1-output Boolean functions. The input specifies $n, s$, and the (partial) Boolean function. The tool returns SAT/UNSAT which means there is/isn't a circuit of size $s$ that can implement the $n$-input 1-output (partial) Boolean function.

More precisely, the problem is given to SAT solver as a SAT formula $\mathsf{lb}_S(f,s)$ of the form $$\bigvee_{a\in S} f(a)\ne C(a)$$ where $S$ is a set of (not necessarily all) $n$-bit strings, $f(a)$ is a value of Boolean function $f$ on input $a$ and $C(a)$ is the 1-output of a circuit $C$ with $s$ gates on input $a$. The circuit $C$ is represented by free variables. The formula $\mathsf{lb}_S(f,s)$ is satisfiable if and only if there exists an $n$-input 1-output circuit $C$ with $s$ gates computing Boolean function $f$ on inputs from the set $S$. Note that the specification of the Boolean function $f$ is hardwired into the SAT formula. 

\def\practical{

\subsection{Work plan}

Step 1. Verify finitistic versions of known circuit lower bounds, e.g. $\PARITY\notin\AC$. The hope is that this could eventually lead to a verification of conjectures such as $\SAT\notin\Ppoly$ or the nonexistence of efficient learning algorithms. It is known \cite{Rkdnf} that Resolution-based SAT solvers cannot prove circuit lower bounds feasibly. Therefore, a new algorithmic approach will be needed. One option is to incorporate a potential speedup provided by the very hardness of circuit lower bounds for Resolution: instead of proving for all, it suffices to prove on the range \cite{}. 
\bigskip

\noindent Step 2. Solve MNIST by SAT solving circuit lower bounds.
\bigskip

\noindent Step 3. Try to break pseudorandom generators of the form $G_C$ from Section \ref{s:leargen}. If the generator survives, its safe against the attack. Otherwise, we get the desired lower bound.
\bigskip

\noindent {\bf Initial challenge.} Prove that DNFs (i.e. depth 2 circuits with OR gates on top and AND gates at the bottom) with 20 inputs and 40 gates cannot compute \PARITY on set $S$ consisting of 2000 randomly chosen 20-bit strings.

\subsection{Encoding}

Naive encoding of circuit lower bounds vs Razborov's encoding.

\subsection{Training samples}

Unsatisfiable instances: take random Boolean functions $f$ and random set of anticheckers. With high probability the resulting lower bound is going to be a tautology.
\smallskip

EF=R* width 3+E (but n literals > $n^3$ clauses > $n^6$ step to find the proof)}
}

\subsection*{Acknowledgements}

I would like to thank Rahul Santhanam for many inspiring discussions which, in particular, motivated me to prove Theorem \ref{t:main}. 
I am indebted to Susanna de Rezende and Erfan Khaniki for many illuminating discussions during the development of the project. I would also like to thank V. Kanade for helpful comments on the existing learning models and L. Chen, V. Kabanets, J. Kraj\'{i}\v{c}ek and I.C. Oliveira for helpful comments on the draft of the paper. This project has received funding from the European Union's Horizon 2020 research and innovation programme under the Marie Sk\l{}odovska-Curie grant agreement No 890220. 
\medskip

\noindent\fbox{\includegraphics[width=48pt,height=31pt]{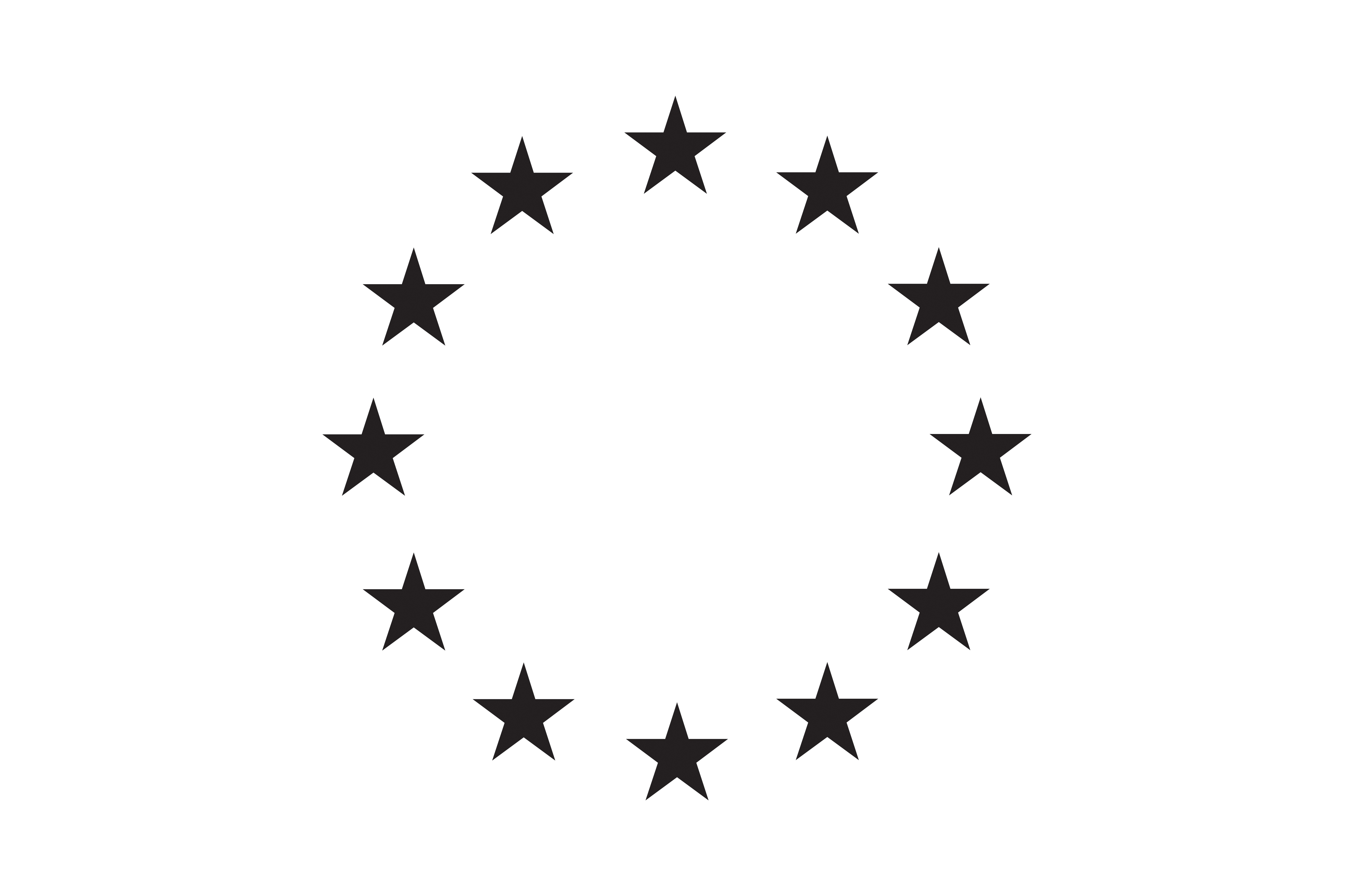}}


\begin{thebibliography}{00}

\bibitem{Alt} Alth\"ofer I.; {\it On sparse approximations to randomized strategies and convex combinations}; Linear Algebra and its Applications, 199(1):339-355, 1994.

\bibitem{Abb} Atserias A.; {\it Distinguishing SAT from polynomial-size circuits, through black-box queries}; CCC, 2006.

\bibitem{BFKL} Blum A., Furst M., Kearns J., Lipton R.; {\it Cryptographic primitives based on hard learning problems}; CRYPTO, 1993.

\bibitem{JOS} Buresh-Oppenheim J., Santhanam R.; {\it Making hard problems harder}; CCC 2006.

\bibitem{CIKK} Carmosino M., Impagliazzo R., Kabanets V., Kolokolova A.; {\it Learning algorithms from natural proofs}; CCC, 2016.

\bibitem{CHOPRS} Chen L., Hirahara S., Oliveira I.C., Pich J., Rajgopal N., Santhanam R.; {\it Beyond natural proofs: hardness magnification and locality}; ITCS, 2020.

\bibitem{CJWs} Chen L., Jin C., Williams R.; {\it Hardness magnification for all sparse \NP languages}; FOCS, 2019.

\bibitem{CJWt} Chen L., Jin C., Williams R.; {\it Sharp threshold results for computational complexity}; STOC, 2020.

\bibitem{CMMW} Chen L., McKay D., Murray C., Williams R.; {\it Relations and equivalences between circuit lower bounds and Karp-Lipton theorems}; CCC, 2019.

\bibitem{CT} Chen L., Tell R.; {\it Bootstrapping results for threshold circuits ``just beyond'' known lower bounds}; STOC, 2019.

\bibitem{CHMY} Cheragchi M., Hirahara S., Myrisiotis D., Yoshida Y.; {\it One-tape Turing machine and read-once branching program lower bounds for MCSP}; preprint, 2020.

\bibitem{GST} Gutfreund D., Shaltiel R., Ta-Shma A.; {\it If NP languages are hard in the worst-case then it is easy to find their hard instances}; CCC, 2005.

\bibitem{Hb} Hirahara S.; {\it Non-black-box worst-case to average-case reductions within \NP}; FOCS, 2018.

\bibitem{Hmeta} Hirahara S.; {\it Non-disjoint promise problems from meta-computational view of pseudorandom generator constructions}; CCC, 2020.

\bibitem{ILO} Ilango R., Loff B., Oliveira I.C.; {\it NP-hardness of circuit minimization for multi-output functions}; CCC, 2020.

\bibitem{Kdual} Kraj\'{i}\v{c}ek J.; {\it Dual weak pigeonhole principle, pseudo-surjective functions and provability of circuit lower bounds}; Journal of Symbolic Logic, 69(1):265-286, 2004.

\bibitem{Knw} Kraj\'{i}\v{c}ek J.; {\it On the proof complexity of the Nisan-Wigderson generator based on a hard $\NP\cap\coNP$ function}; Journal of Symbolic Logic, 11(1):11-27, 2011.

\bibitem{Kfor} Kraj\'{i}\v{c}ek J.; {\it Forcing with random variables and proof complexity}; Cambridge University Press, 2011.

\bibitem{Kht} Kraj\'{i}\v{c}ek J.; {\it On the computational complexity of finding hard tautologies}; Bulletin of the London Mathematical Society, 46(1):111-125, 2014.

\bibitem{Kpc} Kraj\'{i}\v{c}ek J.; {\it Proof complexity}; Cambridge University Press, 2019.

\bibitem{KPT} Kraj\'{i}\v{c}ek J., Pudl\'ak P., Takeuti G.; {\it Bounded arithmetic and the polynomial hierarchy}, Annals of Pure and Applied Logic, 52:143-153, 1991.

\bibitem{LLW} Li L., Littman M., Walsh T.; {\it Knows what it knows: a framework for self-aware learning}; ICML, 2008.

\bibitem{LMN} Linial N., Mansour Y., Nisan N.; {\it Constant depth circuits, Fourier transform, and learnability}; Journal of the Association for Computing Machinery; 40(3):607-620, 1993.

\bibitem{LY} Lipton R.J., Young N.E.; {\it Simple strategies for large zero-sum games with applications to complexity theory}; STOC, 1994.

\bibitem{MMW} McKay D., Murray C., Williams R.; {\it Weak lower bounds on resource-bounded compression imply strong separations of complexity classes}; STOC, 2019.

\bibitem{Msca} Modanese A.; {\it Lower bounds and hardness magnification for sublinear-time shrinking cellular automata}; preprint, 2020.
\bibitem{MP} M\"uller M., Pich J.; 
{\it Feasibly constructive proofs of succinct weak circuit lower bounds}; Annals of Pure and Applied Logic, 2019.  

\bibitem{Nm} Newman I.; {\it Private vs common random bits in communication complexity}; Information Processing Letters, 39:67-71, 1991.

\bibitem{NW} Nisan N., Wigderson A.; {\it Hardness vs. randomness}; J. Comp. Systems Sci., 49:149-167, 1994.

\bibitem{Okol} Oliveira I.C.; {\it Randomness and intractability in Kolmogorov complexity}; ICALP, 2019.

\bibitem{OPS} Oliveira I.C., Pich. J., Santhanam R.; {\it Hardness magnification near state-of-the-art lower bounds}; CCC, 2019.

\bibitem{OS} Oliveira I.C., Santhanam R.; {\it Conspiracies between learning algorithms, circuit lower bounds, and pseudorandomness}; CCC, 2017.

\bibitem{HM} Oliveira I.C., Santhanam R.; {\it Hardness magnification for natural problems}; FOCS, 2018.

\bibitem{Pnw} Pich J.; {\it Nisan-Wigderson generators in proof systems with forms of interpolation}; Mathematical Logic Quarterly, 57(4), 2011.

\bibitem{Pclba} Pich J.; {\it Circuit lower bounds in bounded arithmetics}; Annals of Pure and Applied Logic, 166(1):29-45, 2015.

\bibitem{Pmu15} Pich J.; {\it Mathesis universalis}; Literis, 2016.

\bibitem{PSapc} Pich J., Santhanam R.; {\it Strong co-nondeterministic lower bounds for \NP cannot be proved feasibly}; preprint, 2020.

\bibitem{Rba} Razborov A.A; {\it Unprovability of lower bounds on the circuit size in certain fragments of bounded arithmetic}, Izvestiya of the Russian Academy of Science, 59:201-224, 1995.

\bibitem{Rkdnf} Razborov A.A.; {\it Pseudorandom generators hard for $k$-DNF Resolution and Polynomial Calculus}; Annals of Mathematics, 181(2):415-472, 2015. 

\bibitem{RR} Razborov A.A, Rudich S.; {\it Natural Proofs}; Journal of Computer and System Sciences, 55(1):24-35, 1997.


\bibitem{RS} Rivest R., Sloan R.; {\it Learning complicated concepts reliably and usefully}; AAAI, 1988.

\bibitem{Rs} Rudich S.; {\it Super-bits, demi-bits, and NP/qpoly-natural proofs}; Journal of Computer and System Sciences, 55(1):24-35, 1997.

\bibitem{S19} Santhanam R.; {\it Pseudorandomness and the Minimum Circuit Size Problem}; ITCS, 2020.

\bibitem{Tmag} Tal A.; {\it Computing requires larger formulas than approximating}; STOC, 2017. 

\bibitem{LR} Vadhan S.; {\it Learning versus refutation}; COLT, 2017.

\bibitem{VZ} Vadhan S., Zheng C.J.; {\it A uniform Min-Max theorem with applications in Cryptography}; CRYPTO, 2013.

\end{thebibliography}
\end{document}